# Inductive Reachability Witnesses

ALI ASADI, Sharif University of Technology, Iran
KRISHNENDU CHATTERJEE, IST Austria, Austria
HONGFEI FU, Shanghai Jiao Tong University, China
AMIR KAFSHDAR GOHARSHADY, IST Austria, Austria
MOHAMMAD MAHDAVI, Sharif University of Technology, Iran

In this work, we consider the fundamental problem of reachability analysis over imperative programs with real variables. The reachability property requires that a program can reach certain target states during its execution. Previous works that tackle reachability analysis are either unable to handle programs consisting of general loops (e.g. symbolic execution), or lack completeness guarantees (e.g. abstract interpretation), or are not automated (e.g. incorrectness logic/reverse Hoare logic). In contrast, we propose a novel approach for reachability analysis that can handle general programs, is (semi-)complete, and can be entirely automated for a wide family of programs. Our approach extends techniques from both invariant generation and ranking-function synthesis to reachability analysis through the notion of *(Universal) Inductive Reachability Witnesses* (IRWs/UIRWs). While traditional invariant generation uses over-approximations of reachable states, we consider the natural dual problem of under-approximating the set of program states that can reach a target state. We then apply an argument similar to ranking functions to ensure that all states in our under-approximation can indeed reach the target set in finitely many steps.

On the theoretical level, we first show that our IRW/UIRW-based approach is sound and complete for reachability analysis of imperative programs. Then, we focus on linear and polynomial programs and develop automated methods for synthesizing linear and polynomial IRWs/UIRWs. In the linear case, our techniques are based on Farkas' lemma. For the polynomial case, our approach utilizes three classical theorems in polyhedral geometry and real algebraic geometry, namely Handelman's Theorem, Hilbert's Nullstellensatz and Putinar's Positivstellensatz. To the best of our knowledge, such a combination of these theorems to obtain algorithms for reachability analysis in programs is a novel contribution. On the practical side, our experimental results show that our automated approaches can efficiently prove complex reachability objectives over various benchmarks.

## 1 INTRODUCTION

***Reachability.*** *Reachability analysis* is a basic and fundamental problem in computer science, starting from the halting problem of Turing machines [Turing 1936]. It is a core problem in program verification as it aims at checking whether states with certain properties can be reached during the execution of a program. It also constitutes the most basic liveness property and has been widely studied as a fundamental problem in program analysis and model checking [Clarke et al. 2018; Floyd 1993; Hoare 1969; Manna and Pnueli 2012; Pnueli 1977]. The target states considered in reachablity analysis can be either *desirable* so that reachability to these states should be guaranteed, or *undesirable* so that the goal is to find an execution path leading to an unwanted behavior, hence proving incorrectness of the system. As mentioned, reachability to desirable states encodes the most basic type of liveness property. Reachability to undesirable states is also ubiquitous in verification problems and useful when one needs to identify realistic bugs in software implementations (see e.g. [O'Hearn 2020]). Indeed, in real-world software development, most bugs are identified by finding an execution path that leads to a specific error [Distefano et al. 2019; Godefroid 2007; Majumdar and Sen 2007]. This is the idea that led to developments such as incorrectness logic [O'Hearn 2020].

***Previous Works on Formal Models.*** A large body of research on reachability analysis is conducted over formal models [Clarke et al. 2018], such as finite-state systems [Baier and Katoen 2008, Chapter 3–6], pushdown automata [Walukiewicz 2001], Petri nets [Atig and Ganty 2011; Czerwinski et al. 2019; Darondeau et al. 2012; Mayr 1981] and timed automata [Alur and Dill 1990]. For these models, precise decidability and complexity results are attained. Moreover, numerous efficient algorithms have been developed to automate reachability analysis over these models (see [Baier and Katoen 2008] for a comprehensive overview). Although the formal models above serve as an important abstraction mechanism for realistic systems, the techniques for reachability analysis over thems cannot be applied directly to imperative programs, in particular with real-valued variables. This is because the values taken by the variables in a program typically come from an infinite, even uncountable, domain and the underlying program structure might be irregular, i.e. in many cases a given piece of program code cannot be directly translated into any of the formal models above.

***Reachability in Software Model Checkers.*** Many of the most successful software model checkers rely heavily on reachability analysis [Ball and Rajamani 2002; Beyer et al. 2007; Beyer and Keremoglu 2011; Holzmann 1997]. Notably, the BLAST project [Beyer et al. 2007] describes itself as "a verification tool for the C language that solves the reachability problem". Moreover, even when considering the verification of safety properties, all approaches and tools based on Counterexample-Guided Abstraction Refinement (CEGAR) [Alur et al. 1995; Balarin and Sangiovanni-Vincentelli 1993; Clarke et al. 2000; Gurfinkel et al. 2006; Hajdu and Micskei 2019], including SLAM [Ball et al. 2011; Ball and Rajamani 2002] and BLAST [Beyer et al. 2007], need to constantly perform reachability analyses to obtain their counterexamples. These model checkers rely on predicate abstraction refinement and, assuming that we require the analysis to terminate in finite time, can guarantee completeness when the variables have finite domains [Henzinger et al. 2002], but do not provide such guarantees for programs with real-valued variables.

***Previous Works on (Imperative) Programs.*** When considering imperative programs, the reachability problem, and in particular the special case of termination analysis, has been widely studied over the past decades. There are several relevant categories of previous work, including symbolic execution [Cadar and Sen 2013; Cavada et al. 2014; Jaffar et al. 2012], termination analysis [Floyd 1993], abstract interpretation [Cousot and Cousot 1977] and recent results on incorrectness logic/reverse Hoare logic [de Vries and Koutavas 2011; O'Hearn 2020].

- *Symbolic execution* runs program codes statically in a symbolic fashion, and is thus effective for programs without general unbounded loops. For programs with loops, symbolic execution can only unfold the loop up to a bounded depth, and hence cannot handle general loops with an unbounded number of iterations. This point is also applicable to other approaches that rely on loop unrolling, such as [Albarghouthi et al. 2012a].

- *Termination analysis* is a special kind of reachability that requires the program to reach the terminal program counter, which is usually guaranteed by well-foundedness reasoning such as (lexicographic) ranking functions [Alias et al. 2010; Ben-Amram and Genaim 2015, 2017; Bradley et al. 2005a,b; Colón and Sipma 2001; Cousot 2005; Podelski and Rybalchenko 2004a,b]. Termination analyses do not consider reachability to target program states defined through numerical constraints over program variables.
- *Abstract interpretation* is mainly used to generate over-approximations of reachable states (i.e. certain states *may* be reached), but there are also several abstraction-based approaches that compute under-approximations [Albarghouthi et al. 2012a,b; Giacobazzi et al. 2000; Ranzato 2013; Rival 2005; Schmidt 2007]. However, they cannot provide guarantees of completeness except in specific special cases [Giacobazzi and Ranzato 1997].
- Finally, *incorrectness logic* [O'Hearn 2020] is a sound and complete logic that is similar to Hoare logic but performs under-approximation for reachable program states. A disadvantage of incorrectness logic, much like Hoare logic, is that it requires a considerable amount of manual effort for writing assertions, and cannot be directly automated.

**Previous Works on Invariants.** It is noteworthy that an *invariant* is, in a sense, a dual notion of reachability, and invariant generation is also prominent in the PL literature. Informally, an invariant is an over-approximation of the set of reachable states that can be used to prove safety properties over programs. Invariant generation has been a central research area in program analysis and verification, and many efficient approaches are present, e.g. abstract interpretation [Bagnara et al. 2003; Singh et al. 2017], constraint solving [Chatterjee et al. 2020; Colón et al. 2003; Sankaranarayanan et al. 2004a,b], machine learning [Singh et al. 2018], and abductive inference [Dillig et al. 2013].

**Our Focus.** In this paper, we consider reachability analysis over imperative programs. We study the problem of automatically verifying that a set of target program states can be reached in program execution. While invariants provide an over-approximation of the set of reachable states, we consider their natural dual, i.e. under-approximations of the set of states that can reach a target. We consider programs with non-determinism and distinguish between *existential* and *universal* reachability. *Existential* reachability is the more classical and useful notion and, intuitively speaking, requires that target states are reachable under *some* resolution of the non-deterministic choices in the program. In contrast, *universal* reachability requries the program to reach the target states no matter how the non-determinism is resolved. Our main focus is on existential reachability, but our results generalize to the universal case, as well.

**Our approach.** Our methods are based on constraint solving, and extend ideas from both ranking functions and inductive invariant generation to cover the reachability problem. Informally, we use techniques from inductive invariant generation to capture a subset $\mathbf{T}^\diamond$ of program states from which the execution steps of the program will either reach our target states or stay in $\mathbf{T}^\diamond$ itself. Simultaneously, we use arguments similar to ranking functions to ensure that every state in $\mathbf{T}^\diamond$ can reach a target state in finitely many steps. As mentioned above, the key distinction between our method and invariant generation approaches is that our set $\mathbf{T}^\diamond$ is an *under-approximation* of the set of states that can eventually reach a target state, whereas invariants are, by definition, *over-approximations* of reachable states.

***Our Contributions.*** We propose a novel approach for reachability analysis over programs. In detail, we have the following contributions:
- We propose the novel notion of Inductive Reachability Witnesses (IRWs) for existential reachability, which consists of a state set $\mathbf{T}^{\diamond}$ of program states and a ranking function $f$ over $\mathbf{T}^{\diamond}$. The state set $\mathbf{T}^{\diamond}$ satisfies certain invariant-like conditions. The ranking function $f$ serves as a proof that every state in $\mathbf{T}^{\diamond}$ can indeed reach a target. We also propose the notion of Universal Inductive Reachability Witnesses (UIRWs), the counterpart of IRWs for the universal case.
- From a theoretical point-of-view, we show that IRWs and UIRWs are sound and complete for proving existential and universal reachability, respectively.
- We follow previous template-based works [Chatterjee et al. 2020; Colón et al. 2003; Sankaranarayanan et al. 2004a,b] and use Farkas' Lemma, Putinar's Positivstellensatz, and Handelman's Theorem for automatically synthesizing linear and polynomial IRWs/UIRWs. However, we face new challenges regarding satisfiability in the polynomial case and address them with methods based on Hilbert's Strong Nullstellensatz. To the best of our knowledge, this combination (especially Section 4.3 and Theorem 12) is a novel contribution to constraint-based analysis of polynomial programs. Moreover, it is noteworthy that our synthesis method is complete in the linear case and semi-complete in the polynomial case.
- We show that our completeness results also pay off in practice. We provide experimental results over standard linear benchmarks from SV-COMP 2020 [Beyer 2020]. The results show that in reachability analysis of linear programs, our approach beats every model checker that participated in the competition. Moreover, we present several examples of polynomial programs for which the champions of SV-COMP 2020 fail to prove reachability. In contrast, our approach can successfully handle these programs.

***Novelty.*** Our technical novelty is two-fold. First, we define the sound and complete notions of IRW/UIRW for proving reachability. Second, to capture a large class of imperative programs, we build an automated approach over linear/polynomial transition systems, or equivalently flowchart programs [Alias et al. 2010], where transitions between states with affine/polynomial updates are allowed. We provide (semi-)complete synthesis algorithms based on Farkas' Lemma, Putinar's Positivstellensatz, Handelman's Theorem and Hilbert's Strong Nullstellensatz. While these theorems have previously been used for termination analysis and invariant generation, their application in the context of reachability analysis is novel. Moreover, our combination of Nullstellensätze and Positivstellensätze to obtain program analysis algorithms (see Section 4.3 and Theorem 12) is entirely novel and had not previously been considered even in termination analysis or invariant generation.

## 2 INDUCTIVE REACHABILITY WITNESSES

In this section, we provide the basic definitions needed for reachability analysis, formalize our problems, and introduce the concept of Inductive Reachability Witnesses (IRWs/UIRWs). Finally, we show that IRWs/UIRWs are sound and complete for proving reachability. In the sequel, we use transition systems with real variables to model the programs we are studying.

$I : x \geq 0 \;\wedge\; y \geq 0 \;\wedge\; z \geq 0$

$a:$ **while** $x \geq y:$
$b:$ $\quad (x, y) := (x+1, y+2)$
$c:$ $\quad \square \; (x, y, z) := (x+z, y+z, z-1)$
$d:$

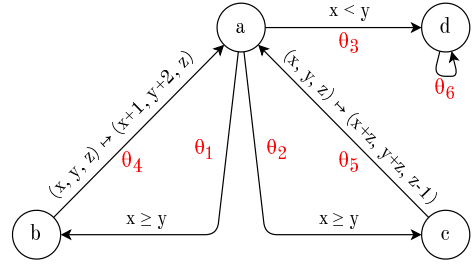

Fig. 1. A Simple Program (left) and its Representation as a Transition System (right)

***Valuations.*** Let $\mathbf{V}$ be a finite set of *variables*. A *valuation* over $\mathbf{V}$ is a function $\nu : \mathbf{V} \to \mathbb{R}$ that assigns a real value to every variable. We denote the set of all valuations over $\mathbf{V}$ by $\mathbb{R}^{\mathbf{V}}$.

***Transition Systems.*** A *transition system* (or simply *system*) is a tuple $S = (\mathbf{V}, \mathbf{L}, \ell_0, I, \Theta)$, in which $\mathbf{V}$ is a finite set of variables, $\mathbf{L}$ is a finite set of *locations*, $\ell_0 \in \mathbf{L}$ is the *initial* or *starting* location, $I$ is an assertion over $\mathbf{V}$ which defines the set of possible initial valuations, and $\Theta$ is a finite set of *transitions*. Each transition $\theta \in \Theta$ is of the form $\theta = (\ell, \ell', \varphi, \mu)$ where $\ell, \ell' \in \mathbf{L}$ are the *pre* and *post* locations, $\varphi$ is an assertion over $\mathbf{V}$ that serves as the *transition condition*, and $\mu : \mathbb{R}^{\mathbf{V}} \to \mathbb{R}^{\mathbf{V}}$ is an *update function*. For brevity, in the sequel, we assume that we have fixed a system $S = (\mathbf{V}, \mathbf{L}, \ell_0, I, \Theta)$ which is under study. For a location $\ell \in \mathbf{L}$, we write $\Theta_\ell$ to denote the set of transitions out of $\ell$. We say that a system is $\beta$-branching if $|\Theta_\ell| \leq \beta$ for every location $\ell$.

**Example 1.** *Consider the program in Figure 1 (left), in which $\square$ denotes non-determinism choice between transitioning to b or c. The transition system in Figure 1 (right) represents this program. Note that we have $\ell_0 = 1$ and assume the initial valuations satisfy $x, y, z \geq 1$.* □

***States.*** A state in $S$ is a pair $\sigma = (\ell, \nu) \in \mathbf{L} \times \mathbb{R}^{\mathbf{V}}$, consisting of a location and a valuation for the variables. We denote the set of all states by $\Sigma$. A subset $\Sigma' \subseteq \Sigma$ of states is called bounded if the set of valuations that appear in the elements of $\Sigma'$ is bounded.

***Successors.*** A state $\sigma' = (\ell', \nu')$ is called a *successor* of a state $\sigma = (\ell, \nu)$ if there exists a transition $\theta = (\ell, \ell', \varphi, \mu) \in \Theta$ such that $\nu \models \varphi$ and $\nu' = \mu(\nu)$. *For theoretical elegance, we assume that every state has at least one successor.* In practice, when modeling a program as a transition system, there might be states in which the program terminates. In such cases, the corresponding transition system will remain in the final state, i.e. we assume that there is a transition from the final state to itself that does not change the value of any variable.

**Example 2.** *In Figure 1 (right), the state $(b, 1, 1, 2)$, i.e. the state at location $b$ for which the values of $x$ and $y$ are $1$ and the value of $z$ is $2$, is a successor of $(a, 1, 1, 2)$ through $\theta_1$. Similarly, $(a, 2, 3, 2)$ is a successor of $(b, 1, 1, 2)$ through $\theta_4$. Also note that there is a transition $\theta_6$ from $d$ to itself, handling the case where the program terminates as described above.* □

***Runs.*** A *run* of the system $S = (\mathbf{V}, \mathbf{L}, \ell_0, I, \Theta)$ is an *infinite* sequence $\mathsf{r} = \{\sigma_i, \theta_i\}_{i=0}^{\infty} = \{(\ell_i, \nu_i), \theta_i\}_{i=0}^{\infty}$, where each $\sigma_i \in \Sigma$ is a state consisting of a location $\ell_i \in \mathbf{L}$ and a valuation $\nu_i \in \mathbb{R}^{\mathbf{V}}$, and each $\theta_i = (\ell_i, \ell_{i+1}, \varphi_i, \mu_i) \in \Theta$ is a transition from $\ell_i$ to $\ell_{i+1}$, such that:
- $\mathsf{r}$ starts in the initial location $\ell_0$;
- $\nu_0 \models I$, i.e. the initial valuation satisfies the assertion $I$;

- For every $i$, we have $\nu_i \models \varphi_i$ and $\nu_{i+1} = \mu_i(\nu_i)$, i.e. $\sigma_{i+1}$ is a successor of $\sigma_i$ through $\theta_i$.

**Semi-runs.** A *semi-run* is defined similarly to a run, except that it does not have to start at $\ell_0$ or satisfy $I$. A *path* of length $n$ is a finite prefix $\pi = \sigma_0, \theta_0, \ldots, \sigma_{n-1}, \theta_{n-1}, \sigma_n$ of a run. Note that a path must always end at a state. Similarly, a *semi-path* is a finite prefix of a semi-run that ends at a state.

**Non-determinism.** The system $S$ is called *deterministic* if there is exactly one possible transition at every state. Formally, $S$ is deterministic if for every $\sigma = (\ell, \nu) \in \Sigma$, there exists exactly one $\theta \in \Theta$ such that $\theta = (\ell, \ell', \varphi, \mu)$ and $\nu \models \varphi$. Otherwise, $S$ is *non-deterministic*.

We are now ready to formalize the reachability problems over transition systems.

**Existential Reachability.** A set $\mathbf{T} \subseteq \Sigma$ is called *existentially reachable* or simply *reachable* if there exists an integer $n$ and a run $\mathbf{r} = \{\sigma_i, \theta_i\}_{i=0}^{\infty}$ such that $\sigma_n \in \mathbf{T}$. In other words, $\mathbf{T}$ is reachable if there exists a run that visits $\mathbf{T}$.

Informally, assuming that $\mathbf{T}$ is a set of undesirable states, e.g. states that lead to a certain error that we would like to avoid, the definition above models the cases when the non-determinism is *demonic* [Back and Wright 2012], i.e. whenever it is possible to choose among multiple transitions, the choice is made in favor of reaching the undesirable set $\mathbf{T}$. However, we can also consider reachability in presence of *angelic* non-determinism [Bodik et al. 2010], i.e. when the choices are made in favor of *not* reaching the undesirable set $\mathbf{T}$. The words "angelic" and "demonic" can have the opposite meaning when considering cases in which $\mathbf{T}$ is a set of desirable states. Therefore, to prevent confusion, we use the terms "existential" and "universal".

**Universal Reachability.** A set $\mathbf{T} \subseteq \Sigma$ is called *universally reachable* if there exists a valuation $\nu_0 \in \mathbb{R}^{\mathbf{V}}$ and an integer $n$, such that (i) $\nu_0 \models I$, and (ii) every run $\mathbf{r} = \{(\ell_i, \nu_i), \theta_i\}_{i=0}^{\infty}$ visits $\mathbf{T}$ in its first $n$ steps. In other words, for each such $\mathbf{r}$, there exists an index $i \leq n$ such that $(\ell_i, \nu_i) \in \mathbf{T}$.

Intuitively, the definition above requires that we can fix an initial valuation for the program such that no matter how the non-determinism is resolved, the execution is forced to visit $\mathbf{T}$ after at most $n$ steps. In this work, our primary focus is on existential reachability. However, our results extend to universal reachability as well.

**Example 3.** *Consider the system in Figure 1 (right), and let $\mathbf{T} = \{(d, \nu) \mid \nu \in \mathbb{R}^{\mathbf{V}}\}$. In this case, reaching $\mathbf{T}$ is equivalent to the termination of the program in Figure 1 (left). Note that $\mathbf{T}$ is existentially reachable, i.e. there are runs of the system that reach label $d$, for example the following:*

$$(a, 0, 0, 0) \xrightarrow{\theta_1} (b, 0, 0, 0) \xrightarrow{\theta_4} (a, 1, 2, 0) \xrightarrow{\theta_3} (d, 1, 2, 0) \to \ldots.$$

*It is also universally reachable, because every execution starting from $(a, 1, 2, 3)$ will reach $d$ in a single step. As another example, consider the target set $\mathbf{T}' = \{(d, \nu) \mid \nu(x) < 0\}$. This corresponds to reaching $d$ (ending the program) with a negative value for $x$. This time, the set $\mathbf{T}'$ is existentially reachable, for example through the following run:*

$$(a, 0, 0, 0) \xrightarrow{\theta_2} (c, 0, 0, 0) \xrightarrow{\theta_5} (a, 0, 0, -1) \xrightarrow{\theta_2} (c, 0, 0, -1) \xrightarrow{\theta_5} (a, -1, -1, -2) \xrightarrow{\theta_2} (c, -1, -1, -2) \xrightarrow{\theta_5}$$
$$(a, -3, -3, -3) \xrightarrow{\theta_1} (b, -3, -3, -3) \xrightarrow{\theta_4} (a, -2, -1, -3) \xrightarrow{\theta_3} (d, -2, -1, -3) \to \ldots,$$

*but it is not universally reachable. To see this, note that if an initial value satisfies $x < y$, then it does not enter the while loop at all, and hence when it reaches $d$ it satisfies $x \geq 0$ (the initial condition). On*

the other hand, if an initial value satisfies $x \geq y$, there is a run that always chooses the transition $\theta_2$ when at $a$, and hence never reaches $T'$. □

We now look into proof concepts for universal and existential reachability.

**T-*inductive Sets*.** Given a set $T \subseteq \Sigma$ of target states, a set $T^\diamond \subseteq \Sigma$ is called T-*inductive* if for every $\sigma \in T^\diamond \setminus T$, there exists a successor $\sigma'$ of $\sigma$ such that $\sigma' \in T^\diamond$.

Intuitively, if $T^\diamond$ is T-inductive, then if we start the execution of the program from a state in $T^\diamond$, there exists a way for resolving the non-determinism so that we either reach $T$ or can inductively prove that we will never leave $T^\diamond$.

**Example 4.** *Consider the system in Figure 1 and let* $T = \{(d, \nu) \mid \nu(x) < 0\}$, *i.e. the target is reaching $d$ with $x$ having a negative value. Let* $T^\diamond := \{(\ell, \nu) \mid \ell \in L, \nu \in \mathbb{R}^V, \nu \models A_\ell\}$ *be the set of states satisfying the following assertions:*

| $\ell$ | $A_\ell$ |
|---|---|
| $a$ | $x, y, z \leq 0 \ \wedge \ (x-y) \cdot (x-y+1) = 0$ |
| $b$ | $x \leq -2 \ \wedge \ y, z \leq 0 \ \wedge \ x = y$ |
| $c$ | $x, y, z \leq 0 \ \wedge \ x = y$ |
| $d$ | $x < 0$ |

*Then, we can verify that $T^\diamond$ is a T-inductive set. Concretely, consider a state $(a, \nu_a) \in T^\diamond$. In other words, $\nu_a \models A_a$. In such a state, we have $(x-y) \cdot (x-y+1) = 0$. Therefore, either $\nu_a(x) = \nu_a(y)$ or $\nu_a(x) = \nu_a(y) - 1$. In the former case, we can take transition $\theta_2$, and it is easy to verify that the new state satisfies $A_c$, hence there is a successor that is also in $T^\diamond$. In the latter case, we can take $\theta_3$ and reach $d$ with a valuation that satisfies $x < 0$, because $\nu_a \models (y \leq 0 \ \wedge \ x = y - 1)$. Similarly, if $(b, \nu_b) \in T^\diamond$, we know that $\nu_b \models (x \leq -2 \ \wedge \ y, z \leq 0 \ \wedge \ x = y)$. Therefore, taking the transition $\theta_4$, corresponding to the update $(x, y) := (x+1, y+2)$, leads to a state in $a$ that satisfies $(x, y, z \leq 0 \ \wedge \ x = y - 1)$. Note that $x = y - 1 \Rightarrow (x-y) \cdot (x-y+1) = 0$, therefore $A_a$ is satisfied and we have a successor in $T^\diamond$. It is easy to verify the same property at $c$. Finally, if we have a state $(d, \nu_d) \in T^\diamond$, by definition of $T$ and $A_d$, we know that $(d, \nu_d) \in T$, and hence we do not need to find any successor for this state.*

*In this example, if we start at an initial state that satisfies $A_a$, we can find a run of the system that either reaches $T$ or stays inside $T^\diamond$. However, this is not enough for reachability to $T$. Such a run might stay inside $T^\diamond$ forever without visiting $T$. For example, we can keep taking the transition $\theta_2$ when at $a$, and hence never reach $d$. To avoid such a scenario, we need a T-ranking function.* □

**T-*ranking Functions*.** Given a T-inductive set $T^\diamond$, a function $f : T^\diamond \to [0, \infty)$ is called a T-*ranking function* with parameter $\epsilon > 0$, if for every $\sigma \in T^\diamond \setminus T$, there exists a successor $\sigma' \in T^\diamond$ of $\sigma$, for which we have $f(\sigma') \leq f(\sigma) - \epsilon$.

***Inductive Reachability Witnesses (IRWs)*.** Given a set $T$ of target states in a system $S = (V, L, \ell_0, I, \Theta)$, an *Inductive Reachability Witness* for $T$ is a tuple $(T^\diamond, f, \epsilon)$ such that:
- $T^\diamond$ is a T-inductive set;
- $\epsilon \in (0, \infty)$;
- $f : T^\diamond \to [0, \infty)$ is a T-ranking function with parameter $\epsilon$;
- There exists a valuation $\nu \in \mathbb{R}^V$ such that $(\ell_0, \nu) \in T^\diamond$ and $\nu \models I$.

Informally, an IRW serves as a proof of existential reachability for a target set $\mathbf{T}$. The inductivity of $\mathbf{T}^\diamond$ ensures that starting from the initial state $(\ell_0, \nu) \in \mathbf{T}^\diamond$, we will never be forced to leave $\mathbf{T}^\diamond$ unless we reach $\mathbf{T}$, while the existence of the $\mathbf{T}$-ranking function $f$ proves that we cannot avoid $\mathbf{T}$ forever. It is also noteworthy that the $\mathbf{T}$-inductive set $\mathbf{T}^\diamond$ is similar to an inductive invariant, but the main difference is that while an invariant is by definition a *superset* of all reachable states, a $\mathbf{T}$-inductive set $\mathbf{T}^\diamond$ is a *subset* of those states from which we can reach the target set $\mathbf{T}$. An IRW $(\mathbf{T}^\diamond, f, \epsilon)$ is called bounded if $\mathbf{T}^\diamond$ is bounded.

**Example 5.** *Consider the system in Figure 1, with the same target set as in Example 4, i.e. $\mathbf{T} = \{(d, \nu) \mid \nu(x) < 0\}$. Let $\mathbf{T}^\diamond := \{(\ell, \nu) \mid \nu \models A_\ell\}$ and $f(\ell, \nu) := f_\ell(\nu)$ be defined as follows:*

| $\ell$ | $A_\ell$ | $f_\ell$ |
|---|---|---|
| $a$ | $-10 \leq x, y, z \leq 0 \wedge \left(x = y - 1 \vee x = y = \frac{-z \cdot (z+1)}{2}\right)$ | $100 + x - y + z$ |
| $b$ | $-10 \leq x \leq -2 \wedge z \leq 0 \wedge x = y = \frac{-z \cdot (z+1)}{2}$ | $99.5 + z$ |
| $c$ | $-2 \leq x \leq 0 \wedge z \leq 0 \wedge x = y = \frac{-z \cdot (z+1)}{2}$ | $99.5 + z$ |
| $d$ | $x \leq -0.5$ | $0$ |

*Note that the $A_\ell$'s are more restrictive than in Example 4. We can verify that $\mathbf{T}^\diamond$ is a $\mathbf{T}$-inductive set in the same manner as in Example 4. We should also verify that $f$ is a valid $\mathbf{T}$-ranking function. Whenever we take either transition $\theta_1$ or $\theta_2$ (from $a$ to $b$ or $c$), we are assured that $x = y$, hence the value of $f$ goes from $100 + z$ to $99.5 + z$ and decreases by $0.5$. Also, because in $A_a$ we have $-10 \leq x, y, z \leq 0$, the value of $f$ at $a$ is at least $80$, and hence transition $\theta_3$ (from $a$ to $d$) decreases $f$ by more than $0.5$. Now consider transition $\theta_4$ (from $b$ to $a$). This transition does not change the value of $z$, but makes it so that $y = x + 1$. So it changes the value of $f$ from $99.5 + z$ to $99 + z$. Note that transition $\theta_5$ (from $c$ to $a$), decreases $z$ by $1$ while keeping $x = y$. Hence, it decreases $f$ by $0.5$. Also, $\theta_6$ (the self-transition from $d$ to $d$) is irrelevant in this case, because our $A_d$ entails inclusion in $\mathbf{T}$. Finally, $(a, 0, 0, 0)$ is a state that satisfies both the initial condition $I$ and $A_a$. Hence, we conclude that $(\mathbf{T}^\diamond, f, 0.5)$ is an IRW for $\mathbf{T}$.* □

**UIRWs.** The definition of a Universal Inductive Reachability Witness (UIRW) is similar to the existential case, except that every part is replaced with its universal counterpart, i.e. the existential quantifiers in $\mathbf{T}^\diamond$ and $f$ are replaced by universal quantifiers. Due to space constraints, we have relegated these definitions to Appendix A.

**Remark 1.** *Note that, as mentioned in Section 1, the inductive set $\mathbf{T}^\diamond$ in Example 5 is an under-approximation of the desired states. In existential IRWs (such as Example 5), the set $\mathbf{T}^\diamond$ is an under-approximation of the states from which there exists a way of resolving the non-determinism so that we eventually reach $\mathbf{T}$. Similarly, in UIRWs (Appendix A), $\mathbf{T}^\diamond$ under-approximates the set of states from which every execution of the program is forced to visit $\mathbf{T}$. Hence, our $\mathbf{T}$-inductive sets $\mathbf{T}^\diamond$ are essentially natural duals of the notion of inductive invariants [Chatterjee et al. 2020; Colón et al. 2003].*

## 3 BASIC RESULTS AND LINEAR/POLYNOMIAL WITNESSES

Our approach for proving existential (resp. universal) reachability is based on synthesizing an IRW (resp. a UIRW). The reduction from reachability to witness synthesis is both sound and complete.

**Theorem 1** (Soundness, Proof in Appendix B). *Let $\mathbf{T} \subseteq \Sigma$ be a set of states in the system $S$.*
  (i) *If there exists an IRW $(\mathbf{T}^\diamond, f, \epsilon)$ for $\mathbf{T}$, then $\mathbf{T}$ is existentially reachable.*

(ii) If there exists a UIRW $(T^\diamond, f, \epsilon)$ for $T$, then $T$ is universally reachable.

**Theorem 2** (Completeness, Proof in Appendix B). *Let $T \subseteq \Sigma$ be a set of states in the system $S$.*
 (i) *If $T$ is existentially reachable, then there exists an IRW $(T^\diamond, f, \epsilon)$ for $T$.*
 (ii) *If $T$ is universally reachable, then there exists a UIRW $(T^\diamond, f, \epsilon)$ for $T$.*

***Undecidability.*** Based on the two theorems above, synthesis of IRWs (UIRWs) is equivalent to proving existential (universal) reachability, which are undecidable problems according to Rice's theorem [Rice 1953]. Hence, whether an arbitrary input system $S$ and target set $T$ have an IRW or a UIRW are undecidable problems, too. As such, in this work we consider the special case of linear or polynomial systems, with target sets that are defined by linear or polynomial inequalities, and focus on the problem of synthesizing linear or polynomial IRWs and UIRWs[*].

***Linear/Polynomial Systems.*** A transition system $S = (V, L, \ell_0, I, \Theta)$ is called $(d, k)$-*polynomial* if
- $I$ is a conjunction of at most $k$ polynomial inequalities of degree at most $d$ over $V$, and
- for every $\theta = (\ell, \ell', \varphi, \mu) \in \Theta$, the transition condition $\varphi$ is a conjunction of at most $k$ polynomial inequalities of degree at most $d$ over $V$, and
- for every $\theta = (\ell, \ell', \varphi, \mu) \in \Theta$ and variable $v \in V$, we have $\mu(v) \in \mathbb{R}[V]$ and $\deg(\mu(v)) \leq d$, i.e. $\mu(v)$ is a polynomial of degree at most $d$ over $V$.

A $(1, k)$-polynomial system is also called $k$-linear.

***Linear IRWs/UIRWs.*** An IRW/UIRW $(T^\diamond, f, \epsilon)$ is called $k$-linear if for every location $\ell \in L$:
- The set $T^\diamond_\ell := T^\diamond \cap (\{\ell\} \times \mathbb{R}^V)$ is a closed polyhedron which is an intersection of at most $k$ half-spaces. In other words, there exists a set $A_\ell$ of at most $k$ non-strict linear inequalities over $V$ such that a valuation $\nu$ satisfies $A_\ell$ iff $(\ell, \nu) \in T^\diamond$.
- The function $f_\ell : \text{Sat}(A_\ell) \to [0, \infty)$, defined as $f_\ell(\nu) = f(\ell, \nu)$, is a linear function over $V$. Here, $\text{Sat}(A_\ell)$ is the set of all valuations that satisfy $A_\ell$.

***Polynomial IRWs/UIRWs.*** An IRW/UIRW $(T^\diamond, f, \epsilon)$ is called $(d, k)$-polynomial if for every $\ell \in L$:
- The set $T^\diamond_\ell := T^\diamond \cap (\{\ell\} \times \mathbb{R}^V)$ is a closed semi-algebraic set defined by at most $k$ non-strict polynomial inequalities of degree $d$ or less. Equivalently, there exists a set $A_\ell$ of at most $k$ non-strict polynomial inequalities of degree at most $d$ over $V$ such that $\nu \models A_\ell$ iff $(\ell, \nu) \in T^\diamond$.
- The function $f_\ell$, defined as $f_\ell(\nu) = f(\ell, \nu)$, is a polynomial of degree at most $d$ over $V$.

A $(d, k)$-polynomial IRW/UIRW is *explicitly bounded* if each set $A_\ell$ contains a polynomial inequality $g \geq 0$ such that $\text{Sat}(g \geq 0)$ is bounded.

## 4 SYNTHESIS OF INDUCTIVE REACHABILITY WITNESSES

We now provide sound and (semi-)complete algorithms for synthesizing linear or polynomial IRWs and UIRWs for linear and polynomial systems. We consider three variants of this problem: (i) when the system, the target set, and the desired IRW/UIRW are all $k$-linear (Section 4.1), (ii) when the system and the desired IRW/UIRW are $k$-linear, but the target set is $(d, k)$-polynomial (Section 4.2), and finally the most general case: (iii) when the system, the target set, and the IRW/UIRW to be synthesized are $(d, k)$-polynomial.

---

[*]All these restrictions are necessary, e.g. termination is undecidable for polynomial programs [Bradley et al. 2005b].

### 4.1 Linear IRWs/UIRWs for Linear Systems with Linear Target Sets

***Problem Definition.*** In this section, we consider the following problem: Given a $k-$linear system $S = (\mathbf{V}, \mathbf{L}, \ell_0, I, \Theta)$, together with a set $\tau_\ell$ of at most $k$ non-strict linear inequalities at every location $\ell \in \mathbf{L}$, synthesize a $k-$linear IRW/UIRW for the target set $\mathbf{T} := \cup_{\ell \in \mathbf{L}} \{\ell\} \times \text{Sat}(\tau_\ell)$ or report that no such IRW/UIRW exists. In the sequel, we assume $\mathbf{V} = \{v_1, \ldots, v_r\}$, and $\mathbf{L} = \{\ell_0, \ldots, \ell_n\}$.

***Mathematical Tool.*** Our approach in this section is based on a well-known theorem in linear programming, called Farkas' Lemma. The presentation we use is similar to [Colón et al. 2003].

**Lemma 1** (Farkas' Lemma [Farkas 1902; Matousek and Gärtner 2007]). *Consider a set $\mathbf{V} = \{v_1, \ldots, v_r\}$ of real-valued variables and the following system of $m$ linear inequalities over $\mathbf{V}$:*

$$\Phi : \begin{cases} a_{1,0} + a_{1,1} \cdot v_1 + \ldots + a_{1,r} \cdot v_r \geq 0 \\ \qquad\qquad\qquad \vdots \\ a_{m,0} + a_{m,1} \cdot v_1 + \ldots + a_{m,r} \cdot v_r \geq 0 \end{cases}$$

*When $\Phi$ is satisfiable, it entails a given linear inequality*

$$\psi : c_0 + c_1 v_1 + \ldots + c_r v_r \geq 0$$

*if and only if $\psi$ can be written as a non-negative linear combination of $1 \geq 0$ and the inequalities in $\Phi$, i.e. if and only if there exist non-negative real numbers $y_0, y_1, \ldots, y_m$, such that:*

$$c_0 = y_0 + \sum_{i=1}^m y_i \cdot a_{i,0} \quad , \quad c_1 = \sum_{i=1}^m y_i \cdot a_{i,1} \quad , \quad \ldots \quad , \quad c_r = \sum_{i=1}^m y_i \cdot a_{i,r}.$$

*Moreover, $\Phi$ is unsatisfiable if and only if $-1 \geq 0$ can be derived as above.*

In our approach, we find ourselves in a situation where $\Phi$ consists of both strict and non-strict linear inequalities. We should hence use the following variant/corollary of Lemma 1:

**Corollary 1** (Proof in Appendix D). *Consider a set $\mathbf{V} = \{v_1, \ldots, v_r\}$ of real-valued variables and the following system of $m$ linear inequalities over $\mathbf{V}$:*

$$\Phi : \begin{cases} a_{1,0} + a_{1,1} \cdot v_1 + \ldots + a_{1,r} \cdot v_r \bowtie_1 0 \\ \qquad\qquad\qquad \vdots \\ a_{m,0} + a_{m,1} \cdot v_1 + \ldots + a_{m,r} \cdot v_r \bowtie_m 0 \end{cases}$$

*in which $\bowtie_i \in \{>, \geq\}$. When $\Phi$ is satisfiable, it entails a given non-strict linear inequality*

$$\psi : c_0 + c_1 v_1 + \ldots + c_r v_r \geq 0$$

*if and only if $\psi$ can be written as a non-negative linear combination of $1 \geq 0$ and the inequalities in $\Phi$, i.e. if and only if there exist non-negative real numbers $y_0, y_1, \ldots, y_m$, such that:*

$$c_0 = y_0 + \sum_{i=1}^m y_i \cdot a_{i,0} \quad , \quad c_1 = \sum_{i=1}^m y_i \cdot a_{i,1} \quad , \quad \ldots \quad , \quad c_r = \sum_{i=1}^m y_i \cdot a_{i,r}.$$

*Moreover, $\Phi$ is unsatisfiable if and only if either $-1 \geq 0$ can be derived as above, or $0 > 0$ can be derived as above with the extra requirement that $\sum_{\bowtie_i \in \{>\}} y_i > 0$ (i.e. in order to derive a strict inequality, we should use at least one of the strict inequalities in $\Phi$ with non-zero coefficient).*

We are now ready to present our algorithm for synthesis of linear IRWs/UIRWs.

***Overview of the Approach.*** Before presenting our algorithm in detail, we provide a high-level overview of its steps. Our algorithm consists of five steps:

- **Step 1.** The algorithm creates a template for the desired IRW/UIRW. Basically, it considers every expression that should be synthesized as part of an IRW/UIRW, i.e. the descriptions of $\mathbf{T}^\diamond$ and $f$, and creates a template for it in which the coefficients are unknown variables whose value should be synthesized.
- **Step 2.** The algorithm generates a series of so-called "constraint pairs". These constraint pairs are of a specific form that is amenable to Farkas' Lemma. They encode the requirements that $\mathbf{T}^\diamond$ should be a $\mathbf{T}$–inductive set and that $f$ should be a valid $\mathbf{T}$-ranking function.
- **Step 3.** In this step, the algorithm applies Farkas' lemma to the constraints generated in Step 2 and translates them to an equivalent system of quadratic (in)equalities over the unknown *template* variables. It is noteworthy that after this step, no program variable appears in the quadratic constraints.
- **Step 4.** The algorithm adds a few additional constraints that ensure the existence of a valid initial valuation for the IRW/UIRW.
- **Step 5.** Finally, the algorithm solves the constraints by calling an off-the-shelf Quadratic Programming (QP) solver. It then plugs back the solution values reported for template variables into the templates generated in Step 1 to obtain the desired IRW/UIRW.

We now dive into the details of each step.

***The Synthesis Algorithm.*** Our algorithm consists of the following five steps:

***Step 1. Setting up a template.*** Consider a $k$–linear IRW/UIRW for reaching $\mathbf{T}$ in $S$. It consists of a $k$–linear set $\mathbf{T}^\diamond$, defined by a set $A_\ell$ of $k$ linear inequalities at every location $\ell$, and a linear function $f$, similarly defined by a linear expression $f_\ell$ at every location $\ell$.

In this step, the algorithm sets up a symbolic *template* for each $A_\ell$ and $f_\ell$. Concretely, it symbolically computes the following expressions, in which the $\widehat{c_{\ell,i,j}}$'s and $\widehat{d_{\ell,j}}$'s are unknown reals[†]:

$$\widehat{A_\ell} : \begin{cases} \widehat{c_{\ell,1,0}} + \widehat{c_{\ell,1,1}} \cdot v_1 + \ldots + \widehat{c_{\ell,1,r}} \cdot v_r \geq 0 \\ \quad\quad\quad\quad\quad\quad \vdots \\ \widehat{c_{\ell,k,0}} + \widehat{c_{\ell,k,1}} \cdot v_1 + \ldots + \widehat{c_{\ell,k,r}} \cdot v_r \geq 0 \end{cases}$$

$$\widehat{f_\ell} = \widehat{d_{\ell,0}} + \widehat{d_{\ell,1}} \cdot v_1 + \ldots + \widehat{d_{\ell,r}} \cdot v_r$$

Intuitively, the goal of the algorithm is to find suitable real values for the unknown coefficients (i.e. $\widehat{c_{\ell,i,j}}$'s and $\widehat{d_{\ell,j}}$'s) so that when we plug them into $\widehat{f_\ell}$'s and $\widehat{A_\ell}$'s, they yield a valid IRW/UIRW. Moreover, the algorithm defines a new unknown $\widehat{\epsilon}$, whose synthesized value will serve as the decrease parameter for $f$.

**Example 6.** *Consider the system in Figure 2. We will use this system as our running example and aim to synthesize a 2-linear IRW and a 2-linear UIRW for it. For the IRW case, suppose that the target* set is $\mathbf{T} = \{(4, \nu) \mid \nu \models (x \geq y + 8)\}$. For the UIRW case, we let $\mathbf{T}' = \{(4, \nu) \mid \nu \models (x \geq y + 4)\}$. In this step, the algorithm generates a variable $\widehat{\epsilon}$ and the following templates:

$$\widehat{A_1} : \begin{cases} \widehat{c_0} + \widehat{c_1} \cdot x + \widehat{c_2} \cdot y \geq 0 \\ \widehat{c_3} + \widehat{c_4} \cdot x + \widehat{c_5} \cdot y \geq 0 \end{cases} \quad \widehat{A_2} : \begin{cases} \widehat{c_6} + \widehat{c_7} \cdot x + \widehat{c_8} \cdot y \geq 0 \\ \widehat{c_9} + \widehat{c_{10}} \cdot x + \widehat{c_{11}} \cdot y \geq 0 \end{cases} \quad \widehat{A_3} : \begin{cases} \widehat{c_{12}} + \widehat{c_{13}} \cdot x + \widehat{c_{14}} \cdot y \geq 0 \\ \widehat{c_{15}} + \widehat{c_{16}} \cdot x + \widehat{c_{17}} \cdot y \geq 0 \end{cases}$$

$$\widehat{A_4} : \begin{cases} \widehat{c_{18}} + \widehat{c_{19}} \cdot x + \widehat{c_{20}} \cdot y \geq 0 \\ \widehat{c_{21}} + \widehat{c_{22}} \cdot x + \widehat{c_{23}} \cdot y \geq 0 \end{cases} \quad \begin{aligned} \widehat{f_1} &= \widehat{d_0} + \widehat{d_1} \cdot x + \widehat{d_2} \cdot y \\ \widehat{f_2} &= \widehat{d_3} + \widehat{d_4} \cdot x + \widehat{d_5} \cdot y \end{aligned} \quad \begin{aligned} \widehat{f_3} &= \widehat{d_6} + \widehat{d_7} \cdot x + \widehat{d_8} \cdot y \\ \widehat{f_4} &= \widehat{d_9} + \widehat{d_{10}} \cdot x + \widehat{d_{11}} \cdot y \end{aligned}$$

---

[†]We use the notation $\widehat{\phantom{x}}$ to denote variables/expressions whose values should be synthesized by the algorithm.

The goal is to synthesize real values for each of the variables $\widehat{\epsilon}, \widehat{c_0}, \ldots, \widehat{c_{23}}$ and $\widehat{d_0}, \ldots, \widehat{d_{11}}$, so that when we plug them back into the templates above, we get a valid IRW/UIRW.

$I : x, y \geq 10$

1 : $if\ x < y:$
2 :     $x := x + 10$
3 :     $\square\ x := x + 5$
4 :

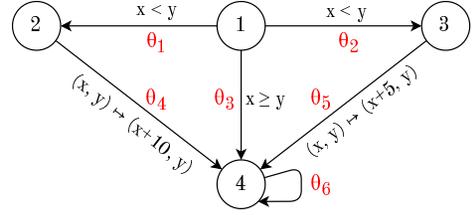

Fig. 2. Our Running Example as a Program (left) and a Transition System (right)

□

**Step 2a. Computing IRW Constraint Pairs.** This step is only performed when we want to synthesize an IRW. In an IRW, the existential inductive set $\mathbf{T}^\diamond$ should satisfy the condition that for every state $\sigma \in \mathbf{T}^\diamond \setminus \mathbf{T}$, there exists a successor $\sigma' \in \mathbf{T}^\diamond$ of $\sigma$. Moreover, there should be at least one such successor for which we have $f(\sigma') \leq f(\sigma) - \epsilon$.

Let $\ell \in \mathbf{L}$ be a location and $\Theta_\ell$ be the set of transitions out of $\ell$, i.e. transitions whose pre-location is $\ell$. The IRW properties at $\ell$ are equivalent to:

$$\forall \nu \in \mathbb{R}^\mathbf{V},\ \nu \models \widehat{A_\ell} \Rightarrow \left( \nu \models \tau_\ell\ \vee\ \bigvee_{\theta = (\ell, \ell', \varphi, \mu) \in \Theta_\ell} \xi(\theta) \right) \quad (1)$$

where $\xi(\theta) = \xi(\ell, \ell', \varphi, \mu)$ is defined as:

$$\xi(\theta) := \left( \nu \models \varphi \wedge \mu(\nu) \models \widehat{A_{\ell'}} \wedge \widehat{f_{\ell'}}(\mu(\nu)) \leq \widehat{f_\ell}(\nu) - \widehat{\epsilon} \right) \quad (2)$$

Intuitively, the constraint in (1) says that if $\nu \models A_\ell$ or equivalently $(\ell, \nu) \in \mathbf{T}^\diamond$, then either $(\ell, \nu) \in \mathbf{T}$ which is equivalent to $\nu \models \tau_\ell$, or there exists a transition $\theta \in \Theta_\ell$, using which we can obtain a successor $(\ell', \mu(\nu)) \in \mathbf{T}^\diamond$ such that $f(\ell', \mu(\nu)) \leq f(\ell, \nu) - \epsilon$. The latter is formalized by $\xi(\theta)$. In this step, the algorithm symbolically computes (1) and writes it in the following equivalent format:

$$\forall \nu \in \mathbb{R}^\mathbf{V},\ \left( \nu \models \widehat{A_\ell} \wedge \bigwedge_{\theta = (\ell, \ell', \varphi, \mu)} \neg \xi(\theta) \right) \Rightarrow \nu \models \tau_\ell \quad (3)$$

Let $P_\ell$ be the LHS assertion in (3) above. Then $P_\ell$ is constructed from logical operations and atomic strict/non-strict linear inequalities over $\mathbf{V}$. Note that the coefficients in these linear inequalities contain the unknown variables $\widehat{c_{\ell,i,j}}$'s and $\widehat{d_{\ell,j}}$'s defined in the previous step.

The algorithm writes $P_\ell$ in disjunctive normal form, obtaining $P_\ell = P_{\ell,1} \vee P_{\ell,2} \vee \ldots \vee P_{\ell,p}$, where each $P_{\ell,i}$ is a conjunction of strict/non-strict linear inequalities over $\mathbf{V}$. It then symbolically computes the following "constraint pair" for every $P_{\ell,i}$:

$$\gamma_{\ell,i} := (P_{\ell,i}, \tau_\ell) \tag{4}$$

The algorithm computes these constraint pairs for every $\ell \in \mathbf{L}$ and stores them in a set $\Gamma$. Note that all computations are symbolic. Every constraint pair $\gamma = (\lambda, \varrho) \in \Gamma$ consists of two parts. $\lambda$ is a set of strict/non-strict linear inequalities, while $\varrho$ is a set of only *non-strict* linear inequalities. Informally, $\gamma$ encodes the requirement that every inequality in $\varrho$ be entailed by inequalities in $\lambda$.

**Example 7.** *Consider the system in Figure 2 together with the templates generated in Example 6. In this step, the algorithm considers location $1 \in \mathbf{L}$, and writes the constraint in (3):*

$$\widehat{A_1} \wedge \neg \xi(\theta_1) \wedge \neg \xi(\theta_2) \wedge \neg \xi(\theta_3) \Rightarrow \tau_1 \tag{5}$$

*Intuitively, the constraint above says that if we are at a $\mathbf{T}^\diamond$ state in location 1 (satisfy $A_1$), and cannot transition to another $\mathbf{T}^\diamond$ with smaller $\widehat{f}$-value using any of the available transitions, in other words $\neg \xi(\theta_1) \wedge \neg \xi(\theta_2) \wedge \neg \xi(\theta_3)$, then we must already be in a target state (satisfy $\tau_1$). There is no target state at location 1, so we can assume $\tau_1 \equiv (-1 \geq 0)$. The algorithm computes (5) symbolically:*

$$\widehat{c_0} + \widehat{c_1} \cdot x + \widehat{c_2} \cdot y \geq 0 \wedge \widehat{c_3} + \widehat{c_4} \cdot x + \widehat{c_5} \cdot y \geq 0 \wedge$$

$$\neg \left( x < y \wedge \widehat{c_6} + \widehat{c_7} \cdot x + \widehat{c_8} \cdot y \geq 0 \wedge \widehat{c_9} + \widehat{c_{10}} \cdot x + \widehat{c_{11}} \cdot y \geq 0 \wedge \widehat{d_3} + \widehat{d_4} \cdot x + \widehat{d_5} \cdot y \leq \widehat{d_0} + \widehat{d_1} \cdot x + \widehat{d_2} \cdot y - \widehat{\epsilon} \right) \wedge$$

$$\neg \left( x < y \wedge \widehat{c_{12}} + \widehat{c_{13}} \cdot x + \widehat{c_{14}} \cdot y \geq 0 \wedge \widehat{c_{15}} + \widehat{c_{16}} \cdot x + \widehat{c_{17}} \cdot y \geq 0 \wedge \widehat{d_6} + \widehat{d_7} \cdot x + \widehat{d_8} \cdot y \leq \widehat{d_0} + \widehat{d_1} \cdot x + \widehat{d_2} \cdot y - \widehat{\epsilon} \right) \wedge$$

$$\neg \left( x \geq y \wedge \widehat{c_{18}} + \widehat{c_{19}} \cdot x + \widehat{c_{20}} \cdot y \geq 0 \wedge \widehat{c_{21}} + \widehat{c_{22}} \cdot x + \widehat{c_{23}} \cdot y \geq 0 \wedge \widehat{d_9} + \widehat{d_{10}} \cdot x + \widehat{d_{11}} \cdot y \leq \widehat{d_0} + \widehat{d_1} \cdot x + \widehat{d_2} \cdot y - \widehat{\epsilon} \right)$$

$$\Rightarrow (-1 \geq 0)$$

*Intuitively, the first line of the constraint above models a state in $\mathbf{T}^\diamond$ at location 1, i.e. it is the same as $\widehat{A_1}$. The second line models the fact that it is not possible to take transition $\theta_1$ and reach another state in $\mathbf{T}^\diamond$ at location 2 such that the $\widehat{f}$-value decreases by at least $\widehat{\epsilon}$. The next two lines model similar constraints for $\theta_2$ and $\theta_3$. Finally, the last line says that if no suitable transition is possible, then the current state must itself be a target, which is impossible in this case because there are no target states at location 1. Next, the algorithm writes the constraint above in disjunctive normal form as:*

$$P_{1,1} \vee P_{1,2} \vee \ldots \vee P_{1,p} \Rightarrow (-1 \geq 0)$$

*Just as before, the algorithm computes each of $P_{1,1}, \ldots, P_{1,p}$ concretely in terms of $x, y, \widehat{\epsilon}, \widehat{c_i}$'s and $\widehat{d_i}$'s, but to save space, we omit the full expansion here. For example, we can assume $P_{1,1}$ is:*

$$\widehat{c_0} + \widehat{c_1} \cdot x + \widehat{c_2} \cdot y \geq 0 \wedge \widehat{c_3} + \widehat{c_4} \cdot x + \widehat{c_5} \cdot y \geq 0 \wedge x \geq y \wedge \widehat{d_9} + \widehat{d_{10}} \cdot x + \widehat{d_{11}} \cdot y > \widehat{d_0} + \widehat{d_1} \cdot x + \widehat{d_2} \cdot y - \widehat{\epsilon}$$

*This corresponds to the case where we cannot use either transition $\theta_1$ or $\theta_2$ because $x \geq y$, and also taking transition $\theta_3$ will lead to a state whose $\widehat{f}$-value is not small enough. For each such $P_{1,i}$ the algorithm generates a constraint pair $(P_{1,i}, \tau_1) = (P_{1,i}, -1 \geq 0)$. The algorithm handles other locations similarly, and adds all the resulting constraint pairs to a set $\Gamma$.* □

***Step 2b. Computing UIRW Constraint Pairs.*** This step is only performed when synthesising a UIRW and is similar to its IRW variant in Step 2a above. Due to space constraints, we have relegated the details of this step to Appendix C.

***Step 2c. Computing Non-negativity Constraints.*** Note than in an IRW/UIRW, the ranking function $f$ should have *non-negative* value over $\mathbf{T}^\diamond$. Let $\ell \in \mathbf{L}$ be a location and $\Theta_\ell$ the set of transitions out of $\ell$. The non-negativity condition at $\ell$ is equivalent to:

$$\forall \nu \in \mathbb{R}^{\mathbf{V}}, \; \nu \models \widehat{A_\ell} \Rightarrow \widehat{f_\ell}(\nu) \geq 0$$

To ensure this constraint, for every $\ell \in \mathbf{L}$, the algorithm adds the constraint pair $(\widehat{A_\ell}, \widehat{f_\ell} \geq 0)$ to $\Gamma$.

**Example 8.** *In the running example, based the templates generated at Example 6, the algorithm creates the following non-negativity constraint pair $\gamma = (\lambda, \varrho)$, encoding $\lambda \Rightarrow \varrho$, at location $1 \in \mathbf{L}$:*

$$\lambda : \begin{cases} \widehat{c_0} + \widehat{c_1} \cdot x + \widehat{c_2} \cdot y \geq 0 \\ \widehat{c_3} + \widehat{c_4} \cdot x + \widehat{c_5} \cdot y \geq 0 \end{cases} \qquad \varrho : (\widehat{d_0} + \widehat{d_1} \cdot x + \widehat{d_2} \cdot y \geq 0)$$

□

***Step 3. Applying Farkas' Lemma.*** The algorithm applies Corollary 1 to every constraint pair generated in the previous step to obtain a non-linear constraint system based on the template variables (i.e. $\widehat{c_{\ell,i,j}}$'s and $\widehat{d_{\ell,j}}$'s), the ranking parameter $\widehat{\epsilon}$, and new variables defined in this step. Crucially, this non-linear constraint system does not include any of the variables in $\mathbf{V}$. We now explain the operations in this step more concretely.

For every constraint pair $\gamma = (\lambda, \varrho) \in \Gamma$, we know that $\lambda$ is a set of strict/non-strict linear inequalities $\{\lambda_{i,0} + \vec{\lambda_i} \cdot \vec{\mathbf{V}} \bowtie_i 0\}_{i=1}^m$, in which $\bowtie_i \in \{>, \geq\}$. Moreover, $\varrho$ is a set of *non-strict* inequalities and every inequality in $\varrho$ should be entailed by $\lambda$. Let $\alpha_0 + \alpha_1 \cdot v_1 + \ldots + \alpha_r \cdot v_r \geq 0 \equiv \alpha_0 + \vec{\alpha} \cdot \vec{\mathbf{V}} \geq 0$ be an inequality in $\varrho$. According to Corollary 1, there are three cases in which $\{\lambda_{i,0} + \lambda_i \cdot \mathbf{V} \bowtie_i 0\}_{i=1}^m$ entails $\alpha_0 + \alpha \cdot \mathbf{V} \geq 0$:

(i) $\alpha_0 + \alpha \cdot \mathbf{V} \geq 0$ is a non-negative combination of $1 \geq 0$ and $\{\lambda_{i,0} + \lambda_i \cdot \mathbf{V} \bowtie_i 0\}_{i=1}^m$, or
(ii) $-1 \geq 0$ can be derived as above, or
(iii) $0 > 0$ can be derived as above.

The algorithm writes constraints that model each of the three cases above and then combines them disjunctively. Given that the three cases are similar, we only explain how (i) is handled: The algorithm creates $m + 1$ new variables $\widehat{y_0}, \widehat{y_1}, \ldots, \widehat{y_m}$ and generates the constraints $\widehat{y_i} \geq 0$ for each one of them. As in Corollary 1, the algorithm computes the following equality symbolically:

$$\alpha_0 + \alpha \cdot \mathbf{V} = \widehat{y_0} + \sum_{i=1}^m \widehat{y_i} \cdot (\lambda_{i,0} + \lambda_i \cdot \mathbf{V}) \qquad (6)$$

Note that the two sides of the equation above are linear expressions over $\mathbf{V}$. As such, they are equal if and only if they agree on the coefficient of every term. The algorithm equates the corresponding coefficients, and adds the following equalities to the constraint system:

$\alpha_0 = \widehat{y_0} + \sum_{i=1}^m \widehat{y_i} \cdot \lambda_{i,0}$     i.e. the constant factor should be equal on both sides

$\forall j \neq 0 \quad \alpha_j = \sum_{i=1}^m \widehat{y_i} \cdot \lambda_{i,j}$     i.e. the coefficient of every variable $v_j \in \mathbf{V}$ should be equal on both sides

The algorithm handles (ii) and (iii) similarly, except that in (iii) we should ensure that at least one strict inequality is used when trying to obtain $0 > 0$. Hence, in this case, the algorithm also adds the extra constraint $\sum_{\bowtie_i \in \{>\}} \widehat{y_i} > 0$ to the non-linear constraint system. The algorithm performs the same operations for every constraint pair $\gamma = (\lambda, \varrho)$ and every linear inequality in $\varrho$ and combines the resulting non-linear constraint systems conjunctively.

**Example 9.** *Consider the constraint pair $\gamma = (\lambda, \varrho)$ below, which was obtained in Example 7:*

$$\lambda : \begin{cases} \widehat{c_0} + \widehat{c_1} \cdot x + \widehat{c_2} \cdot y \geq 0 \\ \widehat{c_3} + \widehat{c_4} \cdot x + \widehat{c_5} \cdot y \geq 0 \\ x - y \geq 0 \\ \widehat{d_9} + \widehat{d_{10}} \cdot x + \widehat{d_{11}} \cdot y - \widehat{d_0} - \widehat{d_1} \cdot x - \widehat{d_2} \cdot y + \widehat{\epsilon} > 0 \end{cases} \qquad \varrho : (-1 \geq 0)$$

*We want to make sure that $\lambda$ entails $\varrho$. Based on Corollary 1, either $\varrho$ or $-1 \geq 0$ or $0 > 0$ should be a non-negative combination of inequalities in $\lambda$. Here, $\varrho$ is itself $-1 \geq 0$, so we only consider two cases:*

- *$-1 \geq 0$ is obtainable from $\lambda$: The algorithm creates 5 new variables $\widehat{y_0}, \widehat{y_1}, \ldots, \widehat{y_4}$ and adds the constraints $\widehat{y_0}, \ldots, \widehat{y_4} \geq 0$. It then computes the following equality:*

$$\widehat{y_0} + \widehat{y_1} \cdot (\widehat{c_0} + \widehat{c_1} \cdot x + \widehat{c_2} \cdot y) + \widehat{y_2} \cdot (\widehat{c_3} + \widehat{c_4} \cdot x + \widehat{c_5} \cdot y) + \widehat{y_3} \cdot (x - y) +$$
$$\widehat{y_4} \cdot (\widehat{d_9} + \widehat{d_{10}} \cdot x + \widehat{d_{11}} \cdot y - \widehat{d_0} - \widehat{d_1} \cdot x - \widehat{d_2} \cdot y + \widehat{\epsilon}) = -1.$$

*Our program variables are $x$ and $y$. All other variables are created by the algorithm and we need to synthesize a value for them. The above is an equality between two polynomials in $\mathbb{R}[x, y]$ that has to hold for all values of $x$ and $y$. Hence, the algorithm equates its corresponding coefficients:*
- $\widehat{y_1} \cdot \widehat{c_1} + \widehat{y_2} \cdot \widehat{c_4} + \widehat{y_3} + \widehat{y_4} \cdot \widehat{d_{10}} - \widehat{y_4} \cdot \widehat{d_1} = 0$ *(the coefficient of $x$ is equal on both sides),*
- $\widehat{y_1} \cdot \widehat{c_2} + \widehat{y_2} \cdot \widehat{c_5} - \widehat{y_3} + \widehat{y_4} \cdot \widehat{d_{11}} - \widehat{y_4} \cdot \widehat{d_2} = 0$ *(the coefficient of $y$ is equal on both sides),*
- $\widehat{y_0} + \widehat{y_1} \cdot \widehat{c_0} + \widehat{y_2} \cdot \widehat{c_3} + \widehat{y_4} \cdot \widehat{d_9} - \widehat{y_4} \cdot \widehat{d_0} + \widehat{y_4} \cdot \widehat{\epsilon} = -1$ *(the constant factor is equal on both sides).*
- *$0 > 0$ is obtainable from $\lambda$: The algorithm creates 5 new variables $\widehat{y_5}, \ldots, \widehat{y_9}$ and proceeds to obtain equalities over non-program variables in the exact same manner as in the previous case, except that it also adds the condition $\widehat{y_9} > 0$.*

□

**Step 4. Computing Initialization Constraints.** By definition, in addition to the inductivity, non-negativity and ranking conditions, an IRW/UIRW should also contain at least one initial state $(\ell_0, \nu)$ such that $\nu \models I$. In other words,

$$\exists \nu_0 = (\nu_{0,1}, \ldots, \nu_{0,r}) \in \mathbb{R}^\mathbf{V}, \nu \models \widehat{A_{\ell_0}} \wedge I. \qquad (7)$$

By $k$−linearity of the system $S$, we know that the initial assertion $I$ is a conjunction of at most $k$ linear inequalities. Thus, the assertion above is a conjunction of at most $2k$ linear inequalities, and is equivalent to $\textsc{Sat}(\widehat{A_{\ell_0}} \wedge I) \neq \emptyset$.

In this step, the algorithm creates $r$ new variables $\widehat{\nu_{0,1}}, \ldots, \widehat{\nu_{0,r}}$, and symbolically computes the linear inequalities in (7), and adds them (conjunctively) to the non-linear constraint system.

**Example 10.** *For our running example (Figure 2), the algorithm creates two new variables $\widehat{\nu_{0,x}}$ and $\widehat{\nu_{0,y}}$ and computes the following:*

$$\begin{array}{ll} \widehat{c_0} + \widehat{c_1} \cdot \widehat{\nu_{0,x}} + \widehat{c_2} \cdot \widehat{\nu_{0,y}} \geq 0 & \widehat{\nu_{0,x}} \geq 10 \\ \widehat{c_3} + \widehat{c_4} \cdot \widehat{\nu_{0,x}} + \widehat{c_5} \cdot \widehat{\nu_{0,y}} \geq 0 & \widehat{\nu_{0,y}} \geq 10 \end{array}$$

*The first two constraints ensure that the valuation $\widehat{\nu_0} = (\widehat{\nu_{0,x}}, \widehat{\nu_{0,y}})$ satisfies $\widehat{A_1}$ and the last two constraints ensure that it satisfies the initial condition I. The algorithm conjunctively adds these constraints to those generated in previous steps.*  □

**Step 5. Solving the Resulting Constraint System.** Finally, the algorithm uses an off-the-shelf solver to solve the resulting non-linear constraint system. If the system is unsatisfiable, it reports that no $k$−linear IRW/UIRW exists. Otherwise, it obtains a solution $\mathfrak{s}$ of the non-linear constraint system. Let $\mathfrak{s}(\widehat{x})$ denote the value assigned by $\mathfrak{s}$ to variable $\widehat{x}$, and extend this definition in the natural way so to any expression $e$. The algorithm outputs $A_\ell := \mathfrak{s}(\widehat{A_\ell})$ and $f_\ell := \mathfrak{s}(\widehat{f_\ell})$, for all $\ell \in L$, as the IRW/UIRW. Moreover, $\mathfrak{s}(\widehat{\nu_{0,1}}, \ldots, \widehat{\nu_{0,r}})$ is the corresponding initial state, and $\mathfrak{s}(\widehat{\epsilon})$ is the decrease parameter for $f$.

**Example 11.** *When the algorithm solves the non-linear (in)equalities obtained in the previous steps, it successfully synthesizes the following IRW[‡] (left table) for $T = \{(4, \nu) \mid \nu \models (x \geq y + 8)\}$ and the following UIRW (right table) for $T' = \{(4, \nu) \mid \nu \models (x \geq y + 4)\}$:*

| $\ell$ | $A_\ell$ | $f_\ell$ |   | $\ell$ | $A_\ell$ | $f_\ell$ |
|---|---|---|---|---|---|---|
| 1 | $y - 2 \leq x \leq y - 1$ | 2 |   | 1 | $y - 0.6 \leq x \leq y - 0.5$ | 2 |
| 2 | $y - 2 \leq x \leq y - 1$ | 1 |   | 2 | $y - 0.6 \leq x \leq y - 0.5$ | 1 |
| 3 | $-1 \geq 0$ | $-1$ |   | 3 | $y - 0.6 \leq x \leq y - 0.5$ | 1 |
| 4 | $x \geq y + 8$ | 0 |   | 4 | $x \geq y + 4.4$ | 0 |

$\epsilon = 1, \quad \nu_0 = (11, 12)$  $\quad\quad$ $\epsilon = 1, \quad \nu_0 = (11, 11.55)$

□

**Theorem 3** (Soundness, Proof in Appendix D). *Given a $k$−linear system $S = (V, L, \ell_0, I, \Theta)$, and a $k$−linear set $T$ of target states, every solution of the non-linear constraint system solved in Step 5 of the algorithm above produces a valid $k$−linear IRW/UIRW for $T$ in $S$.*

**Theorem 4** (Completeness, Proof in Appendix D). *Given a $k$−linear system $S = (V, L, \ell_0, I, \Theta)$, and a $k$−linear set $T$ of target states, every $k$−linear IRW/UIRW for $T$ in $S$ is produced by some solution to the non-linear constraint system solved in Step 5 of the algorithm above.*

**Theorem 5** (Complexity, Proof in Appendix D). *For fixed constants $k$ and $\beta$, given a $k$−linear $\beta$−branching system $S = (V, L, \ell_0, I, \Theta)$, and a $k$−linear set $T$ of target states, Steps 1–4 of the algorithm above lead to a* polynomial-time *reduction from the problem of generating a $k$−linear IRW/UIRW to solving a Quadratic Programming (QP) instance.*

## 4.2 Linear IRWs/UIRWs for Linear Systems with Polynomial Target Sets

In this section, we take the first step towards generalizing our results from the linear case to the polynomial. For technical reasons, we need the concept of strong positivity.

**Strong Positivity.** Let $X \subseteq \mathbb{R}^V$ be a set of valuations and $g \in \mathbb{R}[V]$ a polynomial over $V$. We say that $g$ is *strongly* positive over $X$, and write $X \models g \gg 0$ (or simply $g \gg 0$ when $X$ is clear from context), if $\inf_{x \in X} g(x) > 0$. The real value $\delta := \inf_{x \in X} g(x)$ is called the *positivity gap* of $g$ over $X$. Moreover, $g$ is strongly greater than $h$, denoted $g \gg h$, iff $g - h \gg 0$.

---

[‡]Every solution of the system of non-linear (in)equalities corresponds to a valid IRW. The concrete solution obtained in practice depends on the solver.

***Problem Definition.*** In this section, we consider the following problem: Given a $k$−linear system $S = (\mathbf{V}, \mathbf{L}, \ell_0, I, \Theta)$ together with a set $\tau_\ell$ of at most $k$ *strong* polynomial inequalities of degree at most $d$ at every location $\ell \in \mathbf{L}$, synthesize a $k$−linear IRW/UIRW for a target set $\mathbf{T}$ that satisfies $\tau_\ell$ at every $\ell \in \mathbf{L}$, or report that no such IRW/UIRW exists.

***Mathematical Tool.*** Our main mathematical tool in this section is a theorem, due to Handelman, that characterizes positive polynomials over compact polyhedra. Before presenting this theorem, it is useful to define the notion of monoid.

***Monoid.*** Consider a set $\mathbf{V} = \{v_1, \ldots, v_r\}$ of real-valued variables and the following system of $m$ linear inequalities over $\mathbf{V}$:

$$\Phi : \begin{cases} a_{1,0} + a_{1,1} \cdot v_1 + \ldots + a_{1,r} \cdot v_r \bowtie_1 0 \\ \quad\quad\quad\quad\quad\vdots \\ a_{m,0} + a_{m,1} \cdot v_1 + \ldots + a_{m,r} \cdot v_r \bowtie_m 0 \end{cases}$$

in which $\bowtie_i \in \{>, \geq\}$. Let $g_i$ be the LHS of the $i$-th inequality, i.e. $g_i(v_1, \ldots, v_r) := a_{i,0} + a_{i,1} \cdot v_1 + \ldots + a_{i,r} \cdot v_r$. The monoid of $\Phi$ is defined as:

$$\text{Monoid}(\Phi) := \left\{ \prod_{i=1}^{m} g_i^{\kappa_i} \ \Big| \ \forall 1 \leq i \leq m, \ \kappa_i \in \mathbb{N} \cup \{0\} \right\}.$$

Basically, $\text{Monoid}(\Phi)$ is the set of all polynomials that can be obtained by multiplying the linear expressions on the LHS of $\Phi$ together. Note that each such expression can appear zero or multiple times in the multiplication. Specifically, it is noteworthy that $1 \in \text{Monoid}(\Phi)$.

**Theorem 6** ([Handelman 1988]). *Consider a set $\mathbf{V} = \{v_1, \ldots, v_r\}$ of real-valued variables and the following system of $m$ linear inequalities over $\mathbf{V}$:*

$$\Phi : \begin{cases} a_{1,0} + a_{1,1} \cdot v_1 + \ldots + a_{1,r} \cdot v_r \geq 0 \\ \quad\quad\quad\quad\quad\vdots \\ a_{m,0} + a_{m,1} \cdot v_1 + \ldots + a_{m,r} \cdot v_r \geq 0 \end{cases}$$

*If $\Phi$ is satisfiable, $\text{Sat}(\Phi)$ is compact, and $\Phi$ entails a given polynomial inequality $g(v_1, \ldots, v_r) > 0$ then there exist $y_1, y_2, \ldots, y_u \in [0, \infty)$ and $h_1, h_2, \ldots, h_u \in \text{Monoid}(\Phi)$ such that:*

$$g = \sum_{i=1}^{u} y_i \cdot h_i.$$

As was the case with Farkas' Lemma, it is useful to have a variant of Handelman's theorem that can handle strict inequalities in $\Phi$. We present such a variant, which is a direct corollary of Theorem 6 and characterizes strongly positive polynomials over bounded, but not necessarily closed, polyhedra:

**Corollary 2** (Proof in Appendix D). *Consider a set $\mathbf{V} = \{v_1, \ldots, v_r\}$ of real-valued variables and the following system of $m$ linear inequalities over $\mathbf{V}$:*

$$\Phi : \begin{cases} a_{1,0} + a_{1,1} \cdot v_1 + \ldots + a_{1,r} \cdot v_r \bowtie_1 0 \\ \quad\quad\quad\quad\quad\vdots \\ a_{m,0} + a_{m,1} \cdot v_1 + \ldots + a_{m,r} \cdot v_r \bowtie_m 0 \end{cases}$$

in which $\bowtie_i \in \{>, \geq\}$. If $\Phi$ is satisfiable and $\text{SAT}(\Phi)$ is bounded, then $\Phi$ entails a given strong polynomial inequality $g(v_1, \ldots, v_r) \gg 0$, or in other words $\text{SAT}(\Phi) \models g(v_1, \ldots, v_r) \gg 0$, if and only if there exist constants $y_0 \in (0, \infty)$ and $y_1, y_2, \ldots, y_u \in [0, \infty)$, and polynomials $h_1, h_2, \ldots, h_u \in \text{Monoid}(\Phi)$ such that:

$$g = y_0 + \sum_{i=1}^{u} y_i \cdot h_i. \tag{8}$$

**The Synthesis Algorithm.** Our synthesis algorithm is similar to the one in Section 4.1 and consists of five steps. The main difference is in Step 3, in which constraint pairs are translated to non-linear constraints over template variables. In the previous section, our main tool for this translation was Farkas' Lemma. In this section, due to the more complicated nature of our target sets, we now supplement Farkas' Lemma with Handelman's theorem. For brevity, we do not repeat the presentation of other steps, which are the same as our previous algorithm.

Recall that Step 2 (either Steps 2a and 2c for IRWs, or Steps 2b and 2c for UIRWs) has already generated a set $\Gamma$ of constraint pairs. Each constraint pair $\gamma \in \Gamma$ is of the form $\gamma = (\lambda, \varrho)$ and encodes the requirement that every inequality in $\varrho$ should be entailed by $\lambda$. Moreover, $\lambda$ is a set of strict/non-strict linear inequalities over **V**, whereas $\varrho$ is a set of *strong* polynomial inequalities of degree at most $d$. Let $g \gg 0$ be a strong inequality in $\varrho$. Either $\lambda$ is satisfiable and $g$ should be represented in the form of Equation 8 (cf. Corollary 2) or $\lambda$ is unsatisfiable, in which case $-1 \geq 0$ or $0 > 0$ can be derived as non-negative combinations of inequalities in $\lambda$ and $1 \geq 0$ (cf. Corollary 1).

**Step 3. Applying Handelman's Theorem and Farkas' Lemma.** For every $\gamma = (\lambda, \varrho) \in \Gamma$ and strong polynomial inequality $g \gg 0$ in $\varrho$, the algorithm performs the following operations:
- Let $\text{Monoid}_d(\lambda) = \{h_1, h_2, \ldots, h_u\}$ be the set of all polynomials in $\text{Monoid}(\lambda)$ whose degree is at most $d$. The algorithm symbolically computes $\text{Monoid}_d(\lambda)$ and all of it elements.
- The algorithm considers the following three cases, writes constraints that model each of them, and then combines them disjunctively:
  (i) *Writing $g$ as in Equation 8.* The algorithm creates $u + 1$ new variables $\widehat{y_0}, \widehat{y_1}, \ldots, \widehat{y_u}$ with the constraints $\widehat{y_0} > 0$ and $\widehat{y_1}, \ldots, \widehat{y_u} \geq 0$, and symbolically computes the equation
  
  $$g = \widehat{y_0} + \sum_{i=1}^{u} \widehat{y_i} \cdot h_i.$$
  
  Note that both sides of this equation are polynomials of degree $d$ over **V**. Hence, they are equal iff they agree on the coefficient of every monomial. The algorithm equates the coefficients of corresponding monomials in the LHS and RHS of the equation above, hence obtaining a set of equalities over template variables.
  (ii) *Obtaining $-1 \geq 0$ as a non-negative combination of $\lambda$ and $1 \geq 0$.*
  (iii) *Obtaining $0 > 0$ as a non-negative combination of $\lambda$ and $1 \geq 0$.*
  Cases (ii) and (iii) are handled using Farkas' Lemma in the exact same manner as in our previous algorithm (Section 4.1).
- The algorithm adds the resulting constraints to the non-linear constraint system

**Example 12.** *Consider our running example (Figure 2) together with the templates generated in Example 6. Moreover, assume that we aim to synthesize an IRW for $\tau_3 := (x^2 - x - 100 \gg 0)$, and no target sets in other locations. When Step 2 of the algorithm is applied to location 3 (in exactly the same* *manner as in Section 4.1) it creates several constraint pairs, including the following:*

$$\lambda : \begin{cases} \widehat{c_{12}} + \widehat{c_{13}} \cdot x + \widehat{c_{14}} \cdot y \geq 0 \\ \widehat{c_{15}} + \widehat{c_{16}} \cdot x + \widehat{c_{17}} \cdot y \geq 0 \\ -\widehat{c_3} - 5 \cdot \widehat{c_4} - \widehat{c_4} \cdot x - \widehat{c_5} \cdot y > 0 \end{cases} \qquad \varrho : (x^2 - x - 100 \gg 0)$$

In Step 3 of the algorithm, the constraint pair $\gamma = (\lambda, \varrho)$ is handled as follows:

- The algorithm computes $\text{MONOID}_2(\lambda)$ which consists of all products of polynomials in $\lambda$ up to degree 2. Explicitly, it computes an expanded version of the following polynomials:

$$\begin{aligned}
h_1 &:= 1 & h_2 &:= \widehat{c_{12}} + \widehat{c_{13}} \cdot x + \widehat{c_{14}} \cdot y \\
h_3 &:= \widehat{c_{15}} + \widehat{c_{16}} \cdot x + \widehat{c_{17}} \cdot y & h_4 &:= -\widehat{c_3} - 5 \cdot \widehat{c_4} - \widehat{c_4} \cdot x - \widehat{c_5} \cdot y \\
h_5 &:= h_2^2 & h_6 &:= h_2 \cdot h_3 \\
h_7 &:= h_2 \cdot h_4 & h_8 &:= h_3^2 \\
h_9 &:= h_3 \cdot h_4 & h_{10} &:= h_4^2
\end{aligned}$$

- The algorithm considers cases (i)-(iii) as above. Cases (ii) and (iii) are similar to Section 4.1, so we focus on (i). The algorithm introduces 11 new variables $\widehat{y_0}, \ldots, \widehat{y_{10}}$, adds the constraints $\widehat{y_0} > 0$ and $\widehat{y_1} \ldots \widehat{y_{10}} \geq 0$ and symbolically computes the following equality:

$$x^2 - x - 100 = \widehat{y_0} + \sum_{i=1}^{10} \widehat{y_i} \cdot h_i$$

As before, this is a polynomial equality in $\mathbb{R}[x, y]$, and must hold for all values of $x, y$. So, the corresponding coefficients of the two sides should be equal. The algorithm generates these equalities. For example, given that the constant factor must be the same in the LHS and RHS, the algorithm generates this equality:

$-100 = \widehat{y_0} + \widehat{y_1} + \widehat{y_2} \cdot \widehat{c_{12}} + \widehat{y_3} \cdot \widehat{c_{15}} - \widehat{y_4} \cdot \widehat{c_3} - 5 \cdot \widehat{y_4} \cdot \widehat{c_4} + \widehat{y_5} \cdot \widehat{c_{12}}^2 + \widehat{y_6} \cdot \widehat{c_{12}} \cdot \widehat{c_{15}} - \widehat{y_7} \cdot \widehat{c_{12}} \cdot \widehat{c_3} - 5 \cdot \widehat{y_7} \cdot \widehat{c_{12}} \cdot \widehat{c_4} + \widehat{y_8} \cdot \widehat{c_{15}}^2 - \widehat{y_9} \cdot \widehat{c_{15}} \cdot \widehat{c_3} - 5 \cdot \widehat{y_9} \cdot \widehat{c_{15}} \cdot \widehat{c_4} + \widehat{y_{10}} \cdot \widehat{c_3}^2 + 10 \cdot \widehat{y_{10}} \cdot \widehat{c_3} \cdot \widehat{c_4} + 25 \cdot \widehat{y_{10}} \cdot \widehat{c_4}^2.$

The algorithm generates similar equalities for the coefficients of $x, y, x^2, x \cdot y$, and $y^2$.

□

Note that Steps 4 and 5 are also exactly the same as in our previous algorithm and are omitted here. This being said, we have the following theorems, whose proofs are similar to Section 4.1:

**Theorem 7** (Soundness). *Given a $k$−linear system $S = (V, L, \ell_0, I, \Theta)$, and a set $\tau_\ell$ of at most $k$ polynomial inequalities of degree $d$ or less at every $\ell \in L$, every solution of the non-linear constraint system solved in Step 5 of the algorithm above produces a valid $k$−linear IRW/UIRW for a target set $T$ that satisfies $\tau_\ell$ at every $\ell \in L$.*

**Theorem 8** (Completeness). *Given a $k$−linear system $S = (V, L, \ell_0, I, \Theta)$, and a set $\tau_\ell$ of at most $k$ strong polynomial inequalities of degree $d$ or less at every $\ell \in L$, every bounded $k$−linear IRW/UIRW for a target set $T$ that satisfies $\tau_\ell$ at every $\ell \in L$, is produced by some solution of the non-linear constraint system solved in Step 5 of the algorithm above.*

**Theorem 9** (Complexity). *For fixed constants $k, d$ and $\beta$, given a $k$−linear $\beta$-branching system $S = (V, L, \ell_0, I, \Theta)$, and a set $\tau_\ell$ of at most $k$ polynomial inequalities of degree $d$ or less at every $\ell \in L$, Steps 1–4 of the algorithm above lead to a* polynomial-time *reduction from the problem of generating a $k$-linear IRW/UIRW to solving a QP instance.*

**Remark 2.** *Unlike the linear case, our completeness result in Theorem 8 requires* strong *inequalities and* boundedness. *This is because Handelman's theorem is only applicable when SAT(Φ) is compact, and hence Corollary 2 can only handle strong inequalities over bounded polyhedra. These requirements do not apply to our soundness result, and although they are theoretically necessary, they have very little impact in practice. If there is an IRW/UIRW for a target set* **T** *that ensures reachability within n steps, it is easy to verify that there is also a bounded IRW/UIRW with the same property, i.e. the semi-runs starting at $\nu_0$ and taking n transitions cannot visit an unbounded set of valuations. Moreover, if the target set contains a non-strong inequality such as $g \geq 0$ or $g > 0$, one can replace this inequality with $g + \overline{\epsilon} \gg 0$ for a new variable $\overline{\epsilon} \geq 0$ and solve a quadratic programming instance with the goal of minimizing $\overline{\epsilon}$. This trick will slightly change the problem, but it rarely has practical significance.*

### 4.3 Polynomial IRWs/UIRWs for Polynomial Systems with Polynomial Target Sets

We now provide the most general extension of our algorithm to the case where the system, the target set, and the IRW/UIRW are all polynomial.

***Problem Definition.*** We consider the following problem: Given four technical constants $\Upsilon_1, \ldots, \Upsilon_4 \in \mathbb{N}$, a $(d, k)$–polynomial system $S = (\mathbf{V}, \mathbf{L}, \ell_0, I, \Theta)$, together with a set $\tau_\ell$ of at most $k$ *strong* polynomial inequalities of degree at most $d$ at every location $\ell \in \mathbf{L}$, synthesize a $(d, k)$–polynomial IRW/UIRW for a target set **T** that satisfies $\tau_\ell$ at every $\ell \in \mathbf{L}$, i.e. $\mathbf{T} \cap (\{\ell\} \times \mathbb{R}^\mathbf{V}) \models \tau_\ell$, or report that no such IRW/UIRW exists. The technical constants $\Upsilon_i$ are bounds on the degrees of various polynomials we construct as part of our algorithm. We will soon discuss them more concretely.

***Mathematical Tools.*** We rely on *Putinar's Positivstellensatz* and *Hilbert's Strong Nullstellensatz*. A positivstellensatz (German for positive locus theorem, plural: positvstellensätze) is a theorem in real algebraic geometry that characterizes positive polynomials over semi-algebraic sets. Hilbert's Nullstellensatz (German for zero locus theorem) is a profound theorem that establishes a fundamental relationship between geometry and algebra, and is arguably the basis of the entire field of algebraic geometry. In this work, we use it for solving our satisfiability problems.

**Theorem 10** (Putinar's Positivstellensatz [Putinar 1993]). *Consider a set $\mathbf{V} = \{v_1, \ldots, v_r\}$ of real-valued variables and the following system of m polynomial inequalities over* **V**:

$$\Phi : \left\{ g_1(v_1, \ldots, v_r) \geq 0, \quad \ldots, \quad g_m(v_1, \ldots, v_r) \geq 0 \right.$$

*where $g_1, \ldots, g_m \in \mathbb{R}[\mathbf{V}]$ are polynomials. If there exists a $g_i$ such that the set SAT($g_i \geq 0$) is compact, and $\Phi$ entails a given polynomial inequality $g(v_1, \ldots, v_r) > 0$ then there exist polynomials $h_0, h_1, \ldots, h_m \in \mathbb{R}[\mathbf{V}]$ such that*

$$g = h_0 + \sum_{i=1}^m h_i \cdot g_i$$

*and every $h_i$ is a sum of squares, i.e. $h_i = \sum h_{i,j}^2$ for some polynomials $h_{i,j} \in \mathbb{R}[\mathbf{V}]$.*

Note that the theorem above automatically provides a criterion for satisfiability of Φ. Consider the polynomial inequality $-1 > 0$. This inequality is false, and is hence entailed by Φ if and only if Φ is unsatisfiable. As in the previous sections, we need a variant of this theorem that can handle strict inequalities in Φ. We now obtain such a variant.

**Corollary 3** (Proof in Appendix D). *Consider a set $\mathbf{V} = \{v_1, \ldots, v_r\}$ of real-valued variables and the following system of $m$ polynomial inequalities over $\mathbf{V}$:*

$$\Phi : \left\{ g_1(v_1, \ldots, v_r) \bowtie_1 0, \quad \ldots, \quad g_m(v_1, \ldots, v_r) \bowtie_m 0 \right.$$

*in which every $g_i \in \mathbb{R}[\mathbf{V}]$ is a polynomial and every $\bowtie_i \in \{>, \geq\}$. Also, assume that there is some $i$ such that the set $\textsc{Sat}(g_i \geq 0)$ is compact, or equivalently $\textsc{Sat}(g_i \bowtie_i 0)$ is bounded. If $\Phi$ is satisfiable, then it entails a strong polynomial inequality $g(v_1, \ldots, v_r) \gg 0$, if and only if there exist a constant $y_0 \in (0, \infty)$ and polynomials $h_0, \ldots, h_m \in \mathbb{R}[\mathbf{V}]$ such that*

$$g = y_0 + h_0 + \sum_{i=1}^{m} h_i \cdot g_i \tag{9}$$

*and every $h_i$ is a sum of squares, i.e. $h_i = \sum h_{i,j}^2$ for some polynomials $h_{i,j} \in \mathbb{R}[\mathbf{V}]$.*

Corollary 3 provides a characterization of strongly positive polynomials over $\textsc{Sat}(\Phi)$. However, as in previous sections, we also need a criterion for unsatisfiability of $\Phi$. Given that $\Phi$ may contain both strict and non-strict inequalities, the situation is trickier than Theorem 10. To obtain such a characterization, we use Hilbert's Strong Nullstellensatz for reals.

**Theorem 11** (Strong Nullstellensatz [Atiyah and Macdonald 1969]). *Consider a set $\mathbf{V} = \{v_1, \ldots, v_r\}$ of real-valued variables and let $g_1, \ldots, g_m, g \in \mathbb{R}[\mathbf{V}]$ be polynomials over $\mathbf{V}$. Then exactly one of the following statements holds:*
- *There exists a valuation $\nu \in \mathbb{R}^{\mathbf{V}}$, such that $g_1(\nu) = g_2(\nu) = \ldots = g_m(\nu) = 0$, but $g(\nu) \neq 0$.*
- *There exist a non-negative integer $\alpha$ and polynomials $h_1, \ldots, h_m \in \mathbb{R}[\mathbf{V}]$ such that*

$$\sum_{i=1}^{m} h_i \cdot g_i = g^{\alpha}.$$

We now have the required tools for characterizing unsatisfiable $\Phi$'s.

**Theorem 12** (Proof in Appendix D). *Consider a set $\mathbf{V} = \{v_1, \ldots, v_r\}$ of real-valued variables and the following system of $m$ polynomial inequalities over $\mathbf{V}$:*

$$\Phi : \left\{ g_1(v_1, \ldots, v_r) \bowtie_1 0, \quad \ldots, \quad g_m(v_1, \ldots, v_r) \bowtie_m 0 \right.$$

*in which every $g_i \in \mathbb{R}[\mathbf{V}]$ is a polynomial and every $\bowtie_i \in \{>, \geq\}$. $\Phi$ is unsatisfiable, if and only if at least one of the following statements holds:*

(i) *There exist a constant $y_0 \in (0, \infty)$ and sum-of-square polynomials $h_0, \ldots, h_m \in \mathbb{R}[\mathbf{V}]$ such that*

$$-1 = y_0 + h_0 + \sum_{i=1}^{m} h_i \cdot g_i.$$

(ii) *There exist a non-negative integer $\alpha$ and polynomials $h_1, \ldots, h_m \in \mathbb{R}[\mathbf{V}^*]$ for $\mathbf{V}^* = \mathbf{V} \cup \{w_1, \ldots, w_m\}$, such that for some $1 \leq j \leq m$ with $\bowtie_j \in \{>\}$, we have*

$$w_j^{2 \cdot \alpha} = \sum_{i=1}^{m} h_i \cdot (g_i - w_i^2)$$

**The Synthesis Algorithm.** We are now ready to provide our most general synthesis algorithm for polynomial IRWs/UIRWs over polynomial transition systems. As in the previous cases, our algorithm consists of 5 steps. The main differences are in Steps 1 and 3. In Step 1, our algorithm should now generate a polynomial template. Moreover, in Step 3, it employs Corollary 3 and Theorem 12 for characterizing entailment. The other steps are exactly like our previous algorithms.

**Step 1. Setting up a template.** The algorithm symbolically computes the set of monomials of degree at most $d$ over the variables in $\mathbf{V}$:

$$M_d(\mathbf{V}) := \{\mathfrak{m}_1, \mathfrak{m}_2, \ldots, \mathfrak{m}_u\} := \{v_1^{\alpha_1} \cdot v_2^{\alpha_2} \cdot \ldots \cdot v_r^{\alpha_r} \mid \alpha_1, \ldots, \alpha_r \in \mathbb{N} \cup \{0\} \wedge \alpha_1 + \ldots + \alpha_r \leq d\}.$$

It then sets up the following templates for $A_\ell$ and $f_\ell$ at every location $\ell \in \mathbf{L}$:

$$\widehat{A_\ell} : \begin{cases} \widehat{c_{\ell,1,1}} \cdot \mathfrak{m}_1 + \ldots + \widehat{c_{\ell,1,u}} \cdot \mathfrak{m}_u \geq 0 \\ \qquad\qquad\qquad \vdots \\ \widehat{c_{\ell,k,1}} \cdot \mathfrak{m}_1 + \ldots + \widehat{c_{\ell,k,u}} \cdot \mathfrak{m}_u \geq 0 \end{cases}$$

$$\widehat{f_\ell} = \widehat{d_{\ell,1}} \cdot \mathfrak{m}_1 + \ldots + \widehat{d_{\ell,u}} \cdot \mathfrak{m}_u$$

As usual, the $\widehat{c_{\ell,i,j}}$'s and $\widehat{d_{\ell,j}}$'s are unknown variables for which we should synthesize a value such that the $\widehat{A_\ell}$'s and $\widehat{f_\ell}$'s form an IRW or a UIRW. Note that we do not need to add a separate constant factor to our templates because $1 \in M_d(\mathbf{V})$.

**Step 2. Computing Constraint Pairs.** Steps 2a–2c are the same as in Section 4.1. However, note that the resulting constraint pairs $\gamma = (\lambda, \varrho) \in \Gamma$ are now polynomial. Concretely, $\lambda$ is a set of strict or non-strict polynomial inequalities over $\mathbf{V}$ and $\varrho$ is a set of strong polynomial inequalities over $\mathbf{V}$.

**Step 3. Applying Putinar's Positivstellensatz and Hilbert's Nullstellensatz.** The algorithm applies Corollary 3 and Theorem 12 to every constraint pair generated in the previous step to obtain a non-linear constraint system based on the template variables, $\widehat{\epsilon}$, and new variables defined in this step. Let $\gamma = (\lambda, \varrho) \in \Gamma$ be a constraint pair. $\lambda$ is a set of polynomial inequalities of the form $\{g_i \bowtie_i 0\}_{i=1}^m$. Let $g \gg 0$ be a strong polynomial inequality in $\varrho$. We have to make sure that $\lambda$ entails $g \gg 0$. The algorithm considers three cases:

(i) *$\lambda$ is unsatisfiable due to case (i) in Theorem 12*: The algorithm considers the set $M_{\Upsilon_1}(\mathbf{V}) := \{\mathfrak{m}_1, \mathfrak{m}_2, \ldots, \mathfrak{m}_\mathfrak{n}\}$ of all monomials of degree at most $\Upsilon_1$ over $\mathbf{V}$. Recall that $\Upsilon_1$ is the first technical parameter given in input. It then generates the following templates $\widehat{h_i}$ for $0 \leq i \leq m$:

$$\widehat{h_i} := \widehat{\eta_{i,1}} \cdot \mathfrak{m}_1 + \ldots + \widehat{\eta_{i,\mathfrak{n}}} \cdot \mathfrak{m}_\mathfrak{n}$$

by introducing new variables $\widehat{\eta_{i,j}}$. It also adds certain constraints on $\widehat{\eta_{i,j}}$'s that ensure every $\widehat{h_i}$ is a sum-of-squares. See Appendix E for more details. Then, the algorithm introduces a new variable $\widehat{y_0}$ constrained with $\widehat{y_0} > 0$ and symbolically computes the following equality:

$$-1 = \widehat{y_0} + \widehat{h_0} + \sum_{i=1}^m \widehat{h_i} \cdot g_i$$

Finally, the algorithm equates the corresponding coefficients on the two sides of the equality above, and obtains quadratic equalities over the unknown variables. As before, no program variable appears in these quadratic equalities.

(ii) *$\lambda$ is unsatisfiable due to case (ii) in Theorem 12*: The algorithm considers the set $M_{\Upsilon_2}^* := \{\mathfrak{m}_1^*, \ldots, \mathfrak{m}_{\mathfrak{n}^*}^*\}$ of all monomials of degree at most $\Upsilon_2$ (our second technical parameter) over the extended variable set $\mathbf{V}^* = \mathbf{V} \cup \{w_1, \ldots, w_m\}$. It generates the following templates $\widehat{h_i}$ for $1 \leq i \leq m$:

$$\widehat{h_i} := \widehat{\eta_{i,1}} \cdot \mathfrak{m}_1^* + \ldots + \widehat{\eta_{i,\mathfrak{n}}} \cdot \mathfrak{m}_{\mathfrak{n}^*}^*$$

and symbolically computes the following equality for every index $j$ that corresponds to a *strict* inequality $g_j > 0$ in $\lambda$:

$$w_j^{2 \cdot \Upsilon_3} = \sum_{i=1}^{m} \widehat{h_i} \cdot (g_i - w_i^2).$$

Here $\Upsilon_3$ is our third technical parameter and both sides are polynomials in $\mathbb{R}[\mathbf{V}^*]$. As in the previous case, the algorithm equates the corresponding coefficients on the LHS and RHS and obtains quadratic equalities over unknown variables, i.e. no element of $\mathbf{V}^*$ appears in these equalities. The systems of quadratic equalities generated for each index $j$ are then combined together *disjunctively*.

(iii) *$g$ is a combination of $g_i$'s as in Corollary 3:* The algorithm considers the set $M_{\Upsilon_4} := \{\mathfrak{m}_1, \ldots, \mathfrak{m}_\mathfrak{n}\}$ of monomials of degree at most $\Upsilon_4$ over $\mathbf{V}$, and generates the following templates $\widehat{h_i}$ for $0 \leq i \leq m$:

$$\widehat{h_i} := \widehat{h_i} := \widehat{\eta_{i,1}} \cdot \mathfrak{m}_1 + \ldots + \widehat{\eta_{i,\mathfrak{n}}} \cdot \mathfrak{m}_\mathfrak{n}$$

by introducing new variables $\widehat{\eta_{i,j}}$ and adding constraints that ensure every $\widehat{h_i}$ is a sum-of-squares polynomial (Appendix E). It then introduces a new variable $\widehat{y_0}$ constrained with $\widehat{y_0} > 0$ and symbolically computes this equality:

$$g = \widehat{y_0} + \widehat{h_0} + \sum_{i=1}^{m} \widehat{h_i} \cdot g_i.$$

Finally, the algorithm translates this equality to quadratic equalities over template variables in exactly the same manner as in previous cases.

The systems of quadratic equalities generated in (i)–(iii) above are combined disjunctively.

**Steps 4 and 5.** These steps are exactly the same as those in Section 4.1.

**Example 13.** *Suppose that $\Upsilon_1 = \Upsilon_2 = \Upsilon_3 = \Upsilon_4 = 1$, and the algorithm is in Step 3, handling the following constraint pair:*

$$\lambda : \begin{cases} \widehat{c_1} \cdot x > 0 \\ \widehat{c_2} \cdot y \geq 0 \end{cases} \qquad \varrho : (\widehat{c_3} \cdot x \cdot y + c_4 \gg 0)$$

*The algorithm considers the following cases:*

(i) *It generates three new template polynomials*

$$\widehat{h_0} = \widehat{\eta_{0,1}} + \widehat{\eta_{0,2}} \cdot x + \widehat{\eta_{0,3}} \cdot y$$
$$\widehat{h_1} = \widehat{\eta_{1,1}} + \widehat{\eta_{1,2}} \cdot x + \widehat{\eta_{1,3}} \cdot y$$
$$\widehat{h_2} = \widehat{\eta_{2,1}} + \widehat{\eta_{2,2}} \cdot x + \widehat{\eta_{2,3}} \cdot y$$

*and computes a quadratic system of (in)equalities over the $\widehat{\eta_{i,j}}$'s that ensures every $\widehat{h_i}$ is a sum-of-squares (See Appendix E for details). The algorithm then computes the following equality symbolically (with $\widehat{y_0} > 0$):*

$$-1 = \widehat{y_0} + \widehat{h_0} + \widehat{h_1} \cdot \widehat{c_1} \cdot x + \widehat{h_2} \cdot \widehat{c_2} \cdot y$$

*and rewrites it as quadratic equalities between the unknown variables in the usual way, i.e. by equating the coefficients of corresponding terms on the two sides of the polynomial equality. Intuitively, if there is a valuation for the unknown variables that satisfies these constraints, then $-1$ is a combination of $\widehat{c_1} \cdot x, \widehat{c_2} \cdot y$ and sum-of-square polynomials. Hence, $\lambda$ is unsatisfiable.*

(ii) The algorithm creates two new program variables $w_1, w_2$ and sets up the following templates:

$$\widehat{h_3} = \widehat{\eta_{3,1}} + \widehat{\eta_{3,2}} \cdot x + \widehat{\eta_{3,3}} \cdot y + \widehat{\eta_{3,4}} \cdot w_1 + \widehat{\eta_{3,5}} \cdot w_2$$
$$\widehat{h_4} = \widehat{\eta_{4,1}} + \widehat{\eta_{4,2}} \cdot x + \widehat{\eta_{4,3}} \cdot y + \widehat{\eta_{4,4}} \cdot w_1 + \widehat{\eta_{4,5}} \cdot w_2.$$

Unlike the previous case, $\widehat{h_3}$ and $\widehat{h_4}$ need not be sum-of-squares. It then writes the equality:

$$w_1^2 = \widehat{h_3} \cdot (\widehat{c_1} \cdot x - w_1^2) + \widehat{h_4} \cdot (\widehat{c_2} \cdot y - w_2^2),$$

and converts this polynomial equality to quadratic equalities over the unknown variables by equating the corresponding coefficients. However, note that the LHS and RHS of the polynomial equality above are in $\mathbb{R}[x, y, w_1, w_2]$. According to Theorem 12, any solution to the constraints generated here can serve as a proof for unsatisfiability of $\lambda$.

(iii) The algorithm generates the following template polynomials[§]:

$$\widehat{h_5} = \widehat{\eta_{5,1}} + \widehat{\eta_{5,2}} \cdot x + \widehat{\eta_{5,3}} \cdot y$$
$$\widehat{h_6} = \widehat{\eta_{6,1}} + \widehat{\eta_{6,2}} \cdot x + \widehat{\eta_{6,3}} \cdot y,$$
$$\widehat{h_7} = \widehat{\eta_{7,1}} + \widehat{\eta_{7,2}} \cdot x + \widehat{\eta_{7,3}} \cdot y$$

enforces them to be sum-of-squares just as in case (i) above (Appendix E) and writes the polynomial equality:

$$\widehat{c_3} \cdot x \cdot y + \widehat{c_4} = \widehat{y_1} + \widehat{h_5} + \widehat{h_6} \cdot \widehat{c_1} \cdot x + \widehat{h_7} \cdot \widehat{c_2} \cdot y$$

in which $\widehat{y_1} > 0$. It handles it similarly to the previous cases. Note that this is again a polynomial equality in $\mathbb{R}[x, y]$.

The algorithm combines the systems of quadratic inequality in (i)–(iii) above disjunctively. □

It is now easy to obtain the following theorems, whose proofs are similar to previous cases:

**Theorem 13** (Soundness). *Given a $(d, k)$−polynomial system $S = (\mathbf{V}, \mathbf{L}, \ell_0, I, \Theta)$, and a set $\tau_\ell$ of at most $k$ polynomial inequalities of degree $d$ or less at every $\ell \in \mathbf{L}$, every solution of the non-linear constraint system solved in Step 5 of the algorithm above produces a valid $(d, k)$-polynomial IRW/UIRW for a target set $\mathbf{T}$ that satisfies $\tau_\ell$ at every $\ell \in \mathbf{L}$.*

**Theorem 14** (Semi-completeness). *Consider a $(d, k)$−polynomial system $S = (\mathbf{V}, \mathbf{L}, \ell_0, I, \Theta)$ and a set $\tau_\ell$ of at most $k$ strong polynomial inequalities of degree $d$ or less at every $\ell \in \mathbf{L}$. Let $W = (\mathbf{T}^\diamond, f, \epsilon)$ be an explicitly bounded $(d, k)$−polynomial IRW/UIRW for a target set $\mathbf{T}$ that satisfies $\tau_\ell$ at every $\ell \in \mathbf{L}$. If large enough values are assigned to technical constants $\Upsilon_1, \ldots, \Upsilon_4$, the witness $W$ is produced by some solution of the non-linear constraint system solved in Step 5 of the algorithm above.*

**Theorem 15** (Complexity). *For fixed constants $k, d$ and $\beta$, and technical constants $\Upsilon_1, \ldots, \Upsilon_4$, given a $(k, d)$−polynomial $\beta$−branching system $S = (\mathbf{V}, \mathbf{L}, \ell_0, I, \Theta)$, and a set $\tau_\ell$ of at most $k$ polynomial inequalities of degree $d$ or less at every $\ell \in \mathbf{L}$, Steps 1–4 of the algorithm above lead to a polynomial-time reduction from the problem of generating a $(k, d)$−polynomial IRW/UIRW to solving a QP instance.*

**Remark 3.** *Note that Theorem 14 provides* semi-completeness, *i.e. completeness when the chosen technical constants are large enough. This is because Putinar's Positivstellensatz and Hilbert's Nullstellensatz do not establish a bound on the degree of polynomials that appear in their respective*

---
[§]In practice, the templates in part (i) can be reused for part (iii). This is a simple heuristic that we applied in our implementation and helped decrease the size of the resulting QP.

*characterizations.* Nevertheless, we have to fix a degree in our algorithm when we are generating templates for such polynomials. We use the technical constants $\Upsilon_1, \ldots, \Upsilon_4$ for this purpose. Such semi-completeness results arise routinely in constraint-based termination analysis [Chatterjee et al. 2016] and invariant generation [Feng et al. 2017]. In practice, solutions are often found with small technical constants (see Section 5 for examples).

## 5 EXPERIMENTAL RESULTS

***Implementation.*** We implemented our algorithms for IRW synthesis in Python using SymPy [Meurer et al. 2017] for symbolic computations. The implementation also contains several heuristics for improving performance. Notably, we used Z3 [De Moura and Bjørner 2008] to identify and discard unsatisfiable or tautological constraint pairs, hence reducing the sizes of our QP instances. The QPs were solved by LOQO [Vanderbei 1999]. All results were obtained on an Intel Core i5-2540M (2.6 GHz) machine with 8 GB of RAM running Ubuntu 20.04 LTS. We enforced a time limit of 1800 seconds per verification task.

***Benchmarks.*** For the linear case, we used benchmarks from SV-COMP 2020 [Beyer 2020]. We considered all the tasks in the "Reachability/Safety" category of the competition and removed any benchmarks that asked for safety instead of reachability, or that could not be modeled as transition systems (e.g. due to the presence of pointers or arrays). This left us with 25 benchmarks. For the polynomial case, all standard benchmarks focused on safety. Hence, we created 6 simple programs with complex reachability structure to showcase the strengths of our approach. Due to space restrictions, these examples are put in Appendix F. Specifically, it is noteworthy that these benchmarks demonstrate the fact that our algorithm's success is not dependent on the length or proportion of paths that reach the target set **T**.

***Previous Tools.*** We compare our approach against the two best-performing tools in the Reachability/Safety category of SV-COMP 2020, namely VeriAbs [Afzal et al. 2020] and CPAchecker [Beyer and Keremoglu 2011].

***Linear Results.*** The results over linear benchmarks are summarized in Table 1. Our approach could handle every linear reachability benchmark in SV-COMP 2020. It is noteworthy that according to the SV-COMP results, none of the participating model checkers could handle all 25 benchmarks of Table 1. CPAchecker times out on 9 of the instances, whereas VeriAbs fails on only 1 instance.

After a manual inspection of the benchmarks, we realized that CPAChecker and VeriAbs are faster than our approach when reachability can be attained using liberal abstractions and a relatively short path (benchmarks towards the top of Table 1). This is not surprising, given that in these situations, abstract interpretation and symbolic execution are considerably faster than quadratic constraint solving. However, as the paths to target states become longer and sparser (benchmarks towards the the bottom of Table 1), the advantages of our approach begin to show. When the paths are long, e.g. thousands of steps of program execution, CPAchecker always fails to verify the instance. VeriAbs manages to handle these instances by a clever combination of ideas from loop pruning, loop summarization, abstract interpretation and bounded model checking. However, this comes with a considerable runtime overhead, leading to a much worse performance in comparison with our approach.

| Benchmark | $|L|$ | $|\Theta|$ | $|V|$ | $k$ | $|QP|$ | Gen | Solve | **Ours** | **CPAchecker** | **VeriAbs** |
|---|---|---|---|---|---|---|---|---|---|---|
| gcnr2008 | 8 | 14 | 4 | 2 | 1838 | 14.8 | 81.4 | 96.2 | **1.8** | 17.6 |
| count_up_down-2 | 3 | 4 | 3 | 2 | 244 | 1.6 | 1.9 | **3.5** | 4.4 | 5.9 |
| while_infinite_loop_4 | 10 | 14 | 1 | 2 | 1223 | 5.1 | 7.9 | 13.0 | **4.1** | 15.3 |
| nec11 | 4 | 8 | 3 | 2 | 2871 | 13.2 | 45.7 | 58.9 | **4.2** | 10.8 |
| terminator_02-1 | 5 | 8 | 3 | 3 | 1962 | 12.0 | 19.4 | 31.4 | **4.2** | 17.2 |
| trex02-2 | 5 | 7 | 1 | 2 | 260 | 1.8 | 3.4 | 5.3 | **4.3** | 16.6 |
| multivar_1-2 | 3 | 6 | 2 | 2 | 900 | 5.8 | 19.8 | 25.6 | **4.4** | 9.0 |
| trex01-1 | 14 | 27 | 6 | 3 | 9491 | 69.7 | 228.2 | 297.8 | **4.5** | 17.3 |
| sum04-1 | 6 | 10 | 2 | 2 | 1082 | 5.4 | 8.0 | 13.4 | **5.1** | 17.0 |
| terminator_03-1 | 6 | 11 | 2 | 3 | 1740 | 10.0 | 25.7 | 35.6 | **5.1** | 9.7 |
| trex03-1 | 4 | 12 | 6 | 2 | 8500 | 49.2 | 197.9 | 247.0 | **5.2** | 9.1 |
| for_bounded_loop1 | 10 | 13 | 5 | 2 | 1579 | 9.8 | 30.1 | 39.9 | **5.6** | 16.8 |
| Mono1_1-1 | 3 | 5 | 1 | 2 | 262 | 1.3 | 4.4 | **5.7** | **T/O** | 377.2 |
| sum01_bug02_base | 7 | 13 | 3 | 2 | 7972 | 38.0 | 133.4 | 171.4 | **6.0** | 17.3 |
| sum03-1 | 9 | 14 | 2 | 2 | 20963 | 77.9 | 413.1 | 491.0 | **6.1** | 16.3 |
| id_trans | 5 | 11 | 5 | 2 | 11192 | 68.7 | 171.2 | 239.8 | **6.4** | 19.8 |
| sum01_bug02 | 7 | 12 | 4 | 3 | 17632 | 60.0 | 218.6 | 278.6 | **6.5** | 17.3 |
| sum01-1 | 7 | 12 | 3 | 2 | 7316 | 36.9 | 55.1 | 92.0 | **7.6** | 16.7 |
| nested_1-2 | 4 | 6 | 2 | 2 | 329 | 2.9 | 8.0 | **10.9** | **T/O** | 86.0 |
| const_1-2 | 3 | 6 | 2 | 2 | 901 | 4.6 | 17.6 | **22.2** | **T/O** | 49.6 |
| Mono3_1 | 6 | 8 | 2 | 3 | 660 | 4.0 | 20.0 | **24.0** | **T/O** | 369.9 |
| Mono4_1 | 5 | 7 | 2 | 4 | 949 | 5.3 | 22.1 | **27.3** | **T/O** | 635.8 |
| Mono5_1 | 5 | 7 | 3 | 4 | 1048 | 8.1 | 31.8 | **39.8** | **T/O** | 332.4 |
| Mono6_1 | 5 | 7 | 3 | 5 | 1502 | 11.6 | 48.5 | **60.1** | **T/O** | 382.2 |
| deep_nested | 7 | 17 | 5 | 5 | 3686 | 28.6 | 69.6 | **98.2** | **T/O** | **T/O** |

Table 1. Experimental Results over Linear Reachability Benchmarks from SV-COMP. All times are reported in seconds. "|QP|" is the size of the generated QP instance. "Gen" is the time spent for generating the QP instance and "Solve" is the time spent for solving it. "Ours" is the total runtime of our approach over the instance. The last two columns are the runtimes of CPAchecker and VeriAbs. "T/O" denotes a timeout. The time limit was 1800 seconds per instance. Instances are ordered by the minimum time it took for an approach to solve them.

***Nested Loop Benchmark.*** Figure 3 shows a simple illustration of the main part of deep-nested, the only linear reachability benchmark that could be handled by neither VeriAbs nor CPAchecker. We also ran these tools over this benchmark with an extended time limit of 12 hours, but they timed out. Moreover, according to SV-COMP results, no other participating model checker could handle this example, either. We believe this is because the target state can only be reached after an enormously-long path. Moreover, the target set is quite thin and even the smallest loss of precision in abstraction can cause a failure to prove reachability. However, this particular benchmark is not at all challenging for our method. The runtime of our method does not depend on the length of the paths, and we do not perform abstraction. Moreover, our approach is complete for linear IRWs. As such, it can easily prove reachability in Figure 3.

$$\begin{aligned}
&\textbf{for}\,(a := 0\,;\,a < M - 1\,;\,a := a + 1)\,:\\
&\quad\textbf{for}\,(b := 0\,;\,b < M - 1\,;\,b := b + 1)\,:\\
&\quad\quad\textbf{for}\,(c := 0\,;\,c < M - 1\,;\,c := c + 1)\,:\\
&\quad\quad\quad\textbf{for}\,(d := 0\,;\,d < M - 1\,;\,d := d + 1)\,:\\
&\quad\quad\quad\quad\textbf{for}\,(e := 0\,;\,e < M - 1\,;\,e := e + 1)\,:\\
&\quad\quad\quad\quad\quad\textbf{if}\quad M - 2 \leq a, b, c, d, e\,:\\
&\quad\quad\quad\quad\quad\quad\textbf{print}(\texttt{"target reached"})
\end{aligned}$$

Fig. 3. A simplified version of the deep-nested benchmark. $M > 10^9$ is a very large integer.

| Benchmark | \|L\| | \|Θ\| | \|V\| | $k$ | $d$ | \|QP\| | Gen | Solve | Ours | CPAchecker | VeriAbs |
|---|---|---|---|---|---|---|---|---|---|---|---|
| sqrt2 | 5 | 7 | 2 | 5 | 2 | 2494 | 24.1 | 22.4 | 46.4 | **10.5** | 19.7 |
| sqrt1 | 3 | 4 | 2 | 4 | 2 | 920 | 10.7 | 30.1 | **40.8** | T/O | 207.3 |
| sum | 3 | 4 | 3 | 5 | 2 | 1826 | 20.4 | 59.7 | **80.1** | T/O | F |
| sum2 | 3 | 4 | 3 | 5 | 3 | 2476 | 36.8 | 167.5 | **204.2** | T/O | T/O |
| robot2 | 5 | 8 | 4 | 5 | 2 | 5537 | 71.1 | 681.7 | **752.9** | T/O | F |
| robot1 | 5 | 8 | 4 | 5 | 2 | 5537 | 69.8 | 724.2 | **794.0** | T/O | F |

Table 2. Experimental Results over Polynomial Programs. 'F' denotes that the tool terminated but failed to prove reachability. In all cases, we set our ϒ variables equal to $d$.

***Polynomial Results.*** Table 2 shows our experimental results over 6 polynomial instances. Informally, sqrt1 is a simple program that given an input integer $n \geq 1$ computes $s = \lfloor\sqrt{n}\rfloor$ by trying every possible integer starting from 1. The goal is to (choose a value for $n$ so as to) reach a state with $n - s > 10^5$. sqrt2 is a more clever variant of the same program that doubles the current value in a single step when the doubled value does not exceed $\lfloor\sqrt{n}\rfloor$. sum is a program that sums up all the integers from 1 to $n$. The goal is to synthesize a value for $n$ such that the sum falls in a specific interval. sum2 is a similar benchmark in which the program sums squares of all integers from 1 to $n$. In robot1, two robots are put in the same position in a 2d plane. At each step, each robot non-deterministically chooses to move one unit either upwards or to the right. The goal is to reach a state where the square of the distance between the two robots is more than $10^5$. In robot2, the same two robots are placed on the lower-right and upper-left corners of a square of side length $10^4$. The goal is to show that they can reach a distance of less than 10 from each other. See Appendix F for details.

Similarly to the linear case, we observe that CPAchecker and VeriAbs can handle cases where the path reaching the targets is quite short, and when there is no combinatorial explosion in the number of paths due to repeated non-deterministic choice. Notably, CPAchecker can handle sqrt2 but not sqrt1. The only difference between these two programs is that sqrt2 is more efficient and hence the path to targets is shorter. Moreover, we observe that the various other techniques used by VeriAbs, which made it more successful in the linear case, do not extend well to the polynomial case. In several of the instances, VeriAbs terminates without producing an answer, i.e. reporting unknown as the output. In contrast, our approach is able to handle these examples, given its semi-completeness over polynomial IRWs.

## 6   RELATED WORKS

Below we compare our approach with several families of previous results.

***CEGAR-based Model-Checkers.*** Counterexample-guided abstraction refinement [Abdulla et al. 2016; Alur et al. 1995; Balarin and Sangiovanni-Vincentelli 1993; Clarke et al. 2000] is one of the most successful ideas in software verification and has been implemented in many model-checkers, including the well-known tools SLAM [Ball and Rajamani 2002] and BLAST [Beyer et al. 2007], which handle not only safety properties, but also problems such as test-case generation [Beyer et al. 2004]. These model-checkers repeatedly run reachability analyses in order to obtain the required counterexamples for refining their abstractions. A significant challenge is that when variable domains are infinite or uncountable, e.g. $\mathbb{R}$, they cannot guarantee both termination and completeness. They either provide a complete approach that might not terminate, or a sound terminating approach with no completeness guarantee. Another significant challenge arises when there are many spurious counterexamples. Approaches for mitigating this problem also rely on reachability analysis (e.g. see [Berdine et al. 2013]).

***Invariant Generation.*** Invariant generation aims to automatically generate over-approximations of reachable sets, while our approach targets reachability analysis that aims to check whether certain undesirable states can be reached (whether a bug is present in the program). Note that although we have the same inductiveness idea as in inductive invariant generation, the idea leads to *under-approximation* (rather than over-approximation) of the set of states that can reach some target state. Invariant generation is very well-studied and many approaches are developed for automating it, including recurrence analysis [Farzan and Kincaid 2015; Kincaid et al. 2017, 2018], abstract interpretation [Bagnara et al. 2005; Chakarov and Sankaranarayanan 2014; Feautrier and Gonnord 2010; Gawlitza and Monniaux 2012], constraint-solving [Chatterjee et al. 2020; Colón et al. 2003], inference [Sharma and Aiken 2014] and symbolic execution [Csallner et al. 2008].

***Symbolic Execution.*** Symbolic execution [Burch et al. 1990; Cadar and Sen 2013] runs program codes in a static and symbolic manner, thus it is very effective for analyzing programs without loops or with bounded loops. When tackling loops, symbolic execution can only unfold the loop a bounded number of steps, hence it is unsuitable for loops with an unbounded number of iterations. In contrast, our approach can handle loops with unbounded iterations, and provides soundness and completeness guarantees.

***Abstract Interpretation.*** Abstract interpretation mainly focuses on over-approximation of reachable states through the widening operator, which often leads to a loss of precision [Gonnord and Schrammel 2014]. There are also several abstraction-based results on under-approximation [Giacobazzi et al. 2000; Ranzato 2013; Rival 2005; Schmidt 2007]. However, a theory with completeness guarantees for generating under-approximations such as our T-inductive sets through abstract interpretation is still lacking.

***Termination Analysis.*** Termination analysis only considers whether a program terminates in a finite number of steps, i.e. whether the program can reach the terminal program counter or not, without constraints over program variables. In our approach, we consider reachability to program states defined by numerical constraints over program variables, which is a considerable extension of the termination property. The primary method of proving termination is to synthesize a ranking function, and there are template-based algorithms for the synthesis of linear/polynomial ranking functions [Chatterjee et al. 2016; Colón and Sipma 2001; Leike and Heizmann 2014; Podelski and Rybalchenko 2004a]. Termination and reachability properties have also been extensively studied in the context of probabilistic programs, especially through martingale-based approaches (e.g. see [Chakarov and Sankaranarayanan 2013; Huang et al. 2019; Takisaka et al. 2018]).

**Incorrectness Logic.** Incorrectness logic [O'Hearn 2020] and reverse Hoare logic [de Vries and Koutavas 2011] are similar to Hoare logic but target under-approximations. The logical background behind our approach is a bit different from incorrectness logic. Incorrectness logic obtains under-approximations of the set of reachable states. Hence, a bug can be found by taking the intersection of the under-approximation obtained by incorrectness logic and the set **T** of undesirable states. In contrast, we find under-approximations of the sets of states from which reachability to an undesirable state (or bug) in **T** is guaranteed. Intuitively, the relationship between our approach and incorrectness logic is similar to the relationship between inductive invariants and Hoare logic. It is also noteworthy that incorrectness logic needs manual effort to write assertions, while our approach is entirely automated when we consider linear/polynomial IRWs/UIRWs.

## 7 CONCLUSION

In this work, we proposed the new approach of inductive reachability witnesses for reachability analysis over imperative programs. Our approach extends methodologies from both ranking-function synthesis and invariant generation to tackle reachability. It synthesizes an under-approximation of the set of program states that reach some target state, then uses a ranking argument to ensure eventual reachability. On the theoretical side, we proved that our approach is sound and complete when there is no restriction over the form of inductive reachability witnesses, and presented automated sound and (semi-)complete algorithms for synthesizing linear and polynomial inductive reachability witnesses. On the practical side, our experimental results show that our automated approaches are applicable and can efficiently solve non-trivial reachability objectives over various complex benchmarks.

An interesting future direction would be to incorporate more advanced ranking-function synthesis methods such as lexicographic ranking functions [Alias et al. 2010; Ben-Amram and Genaim 2017; Bradley et al. 2005a] into reachability analysis. Another possible future direction is to consider how our approach can be extended to automate the search for proofs in incorrectness logic [O'Hearn 2020]. Inductive invariant generation has been successfully used as a prerequisite in automating some aspects of Hoare logic and termination analysis (e.g. see [Chatterjee et al. 2016; Colón and Sipma 2001; Podelski and Rybalchenko 2004a]). Given that our witnesses are natural duals of inductive invariants in the same manner that incorrectness logic is dualizing Hoare's logic, we expect that this direction will be fruitful.

30   Ali Asadi, Krishnendu Chatterjee, Hongfei Fu, Amir Kafshdar Goharshady, and Mohammad Mahdavi

## A UNIVERSAL REACHABILITY WITNESSES

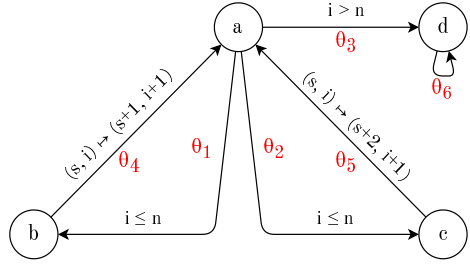

$I : i = s = 0 \wedge n \geq 0$

$a:$ **while** $i \leq n$:
$b:$ $\quad (s, i) := (s + 1, i + 1)$
$c:$ $\quad \square \ (s, i) := (s + 2, i + 1)$
$d:$

Fig. 4. A Non-deterministic Program (left) and its Representation as a Transition System (right)

In this section, we provide detailed definitions and an example of the Universal Inductive Reachability Witnesses (UIRWs).

**Universal T-inductive Sets.** Given a set $T \subseteq \Sigma$ of target states, a set $T^\diamond \subseteq \Sigma$ is called *universally* T-*inductive* if for every $\sigma \in T^\diamond \setminus T$ and *every* successor $\sigma'$ of $\sigma$, we also have $\sigma' \in T^\diamond$.

The idea behind universal T-inductive sets is that any execution of the program that starts in such a set $T^\diamond$ will either reach $T$ or one can prove using induction that it will never leave $T^\diamond$, no matter how the non-determinism is resolved.

**Universal T-ranking Functions.** Given a universal T-inductive set $T^\diamond$, a function $f : T^\diamond \to [0, \infty)$ is called a *universal* T-*ranking function* with parameter $\epsilon > 0$, if for every $\sigma \in T^\diamond \setminus T$ and *every* successor $\sigma'$ of $\sigma$, we have $f(\sigma') \leq f(\sigma) - \epsilon$.

**Universal Inductive Reachability Witnesses (UIRWs).** Given a set $T$ of target states in a system $S = (V, L, \ell_0, I, \Theta)$, a *Universal Inductive Reachability Witness* for $T$ is a tuple $(T^\diamond, f, \epsilon)$ such that:
- $T^\diamond$ is a universal T-inductive set;
- $\epsilon \in (0, \infty)$;
- $f : T^\diamond \to [0, \infty)$ is a universal T-ranking function with parameter $\epsilon$;
- There exists a valuation $\nu \in \mathbb{R}^V$ such that $(\ell_0, \nu) \in T^\diamond$ and $\nu \models I$.

**Example 14.** *Figure 4 shows a simple program together with its representation as a transition system. Let* $T = \{(d, \nu) \mid \nu(s) \geq 20\}$, *i.e. the target is reaching point $d$ with an $s$ value of more than 20. Let* $T^\diamond := \{(\ell, \nu) \mid \nu \models A_\ell\}$ *and* $f(\ell, \nu) := f_\ell(\nu)$ *be defined as follows:*

| $\ell$ | $A_\ell$ | $f_\ell$ |
| --- | --- | --- |
| $a$ | $n \geq 50 \wedge s \geq i \geq 0 \wedge n + 1 \geq i$ | $n + 1.5 - i$ |
| $b$ | $n \geq 50 \wedge s, n \geq i \geq 0$ | $n + 1 - i$ |
| $c$ | $n \geq 50 \wedge s, n \geq i \geq 0$ | $n + 1 - i$ |
| $d$ | $s \geq 50$ | $0$ |

*It is easy to check that* $(T^\diamond, f, 0.5)$ *is a UIRW for* $T$. *Intuitively, this guarantees that if a run starts with an initial valuation that satisfies $A_a$, it will definitely reach a target state.* □

## B PROOFS OF THEOREMS PRESENTED IN SECTION 3

In this section we provide proofs of our basic soundness and completeness theorems.

**Theorem 1** (Soundness). Let $\mathbf{T} \subseteq \Sigma$ be a set of states in the system $S = (\mathbf{V}, \mathbf{L}, \ell_0, I, \Theta)$.

(i) If there exists an IRW $(\mathbf{T}^\diamond, f, \epsilon)$ for $\mathbf{T}$, then $\mathbf{T}$ is existentially reachable.
(ii) If there exists a UIRW $(\mathbf{T}^\diamond, f, \epsilon)$ for $\mathbf{T}$, then $\mathbf{T}$ is universally reachable.

PROOF. We handle each case separately.

(i) We construct a run of $S$ that visits $\mathbf{T}$. By definition of IRW, there exists a state $\sigma_0 = (\ell_0, \nu_0) \in \mathbf{T}^\diamond$ such that $\nu_0 \models I$. We start our run with $\sigma_0$ and inductively find the next transitions and states as follows: when we are in a state $\sigma_i \in \mathbf{T}^\diamond$, either (a) $\sigma_i \in \mathbf{T}$ in which case the path until this point has already reached $\mathbf{T}$ and we can extend it to an arbitrary run, or (b) $\sigma_i \in \mathbf{T}^\diamond \setminus \mathbf{T}$, in which case there exists a successor $\sigma_{i+1} \in \mathbf{T}^\diamond$ of $\sigma_i$ such that $f(\sigma_{i+1}) \leq f(\sigma_i) - \epsilon$, and we transition to $\sigma_{i+1}$. Using this procedure, it is not possible to avoid case (a) forever, because each application of (b) decreases the value of $f$ by at least $\epsilon$ and $f$ is bounded from below. Hence, the constructed run will reach $\mathbf{T}$.

(ii) We choose $\sigma_0 = (\ell_0, \nu_0)$ as in the previous case. We now prove that every path of length $n := 1 + \lceil f(\sigma_0)/\epsilon \rceil$ starting from $\sigma_0$ will reach $\mathbf{T}$. Let $r = \sigma_0, \theta_0, \sigma_1, \theta_1, \ldots, \sigma_n$ be such a path. If no $\sigma_i$ is in $\mathbf{T}$, then by definition of universal $\mathbf{T}$-inductiveness, every $\sigma_i$ is in $\mathbf{T}^\diamond \setminus \mathbf{T}$. So, for each $i$, we have $f(\sigma_{i+1}) \leq f(\sigma_i) - \epsilon$. Therefore, $f(\sigma_n) \leq f(\sigma_0) - n \cdot \epsilon = f(\sigma_0) - \epsilon - \lceil f(\sigma_0)/\epsilon \rceil \cdot \epsilon < 0$ which is a contradiction because $f$ can only take non-negative values.

□

**Theorem 2** (Completeness). Let $\mathbf{T} \subseteq \Sigma$ be a set of states in the system $S = (\mathbf{V}, \mathbf{L}, \ell_0, I, \Theta)$.

(i) If $\mathbf{T}$ is existentially reachable, then there exists an IRW $(\mathbf{T}^\diamond, f, \epsilon)$ for $\mathbf{T}$.
(ii) If $\mathbf{T}$ is universally reachable, then there exists a UIRW $(\mathbf{T}^\diamond, f, \epsilon)$ for $\mathbf{T}$.

PROOF. In each case, we construct the required IRW/UIRW.

(i) Given that $\mathbf{T}$ is reachable, by definition there exists a path $\pi = (\ell_0, \nu_0), \theta_0, \ldots, (\ell_n, \nu_n)$ such that $(\ell_n, \nu_n) \in \mathbf{T}$ and $\nu_0 \models I$. Without loss of generality, we choose such a $\pi$ that is prefix-minimal, i.e. that no prefix of $\pi$ has the same properties. Let $\mathbf{T}^\diamond = \{(\ell_i, \nu_i) \mid 0 \leq i \leq n\}$, then $\mathbf{T}^\diamond$ is $\mathbf{T}$-inductive, because $(\ell_n, \nu_n) \in \mathbf{T}$ and for every $i \neq n$, the state $(\ell_i, \nu_i)$ can be succeeded by $(\ell_{i+1}, \nu_{i+1})$. Let $f : \mathbf{T}^\diamond \to [0, \infty)$ be defined as follows: $f(\ell_i, \nu_i) := n - i$. It is easy to verify that $(\mathbf{T}^\diamond, f, 1)$ is an IRW for $\mathbf{T}$.

(ii) We define $\Sigma_k \subseteq \Sigma$ as the set of all states such that every semi-path of length $k$ starting in these states is guaranteed to visit $\mathbf{T}$. Note that $\Sigma_0 = \mathbf{T}$ and if $\sigma \in \Sigma_k \setminus \mathbf{T}$, then by definition every successor $\sigma'$ of $\sigma$ must be in $\Sigma_{k-1}$. Let $\mathbf{T}^\diamond = \bigcup_{i=0}^\infty \Sigma_k$, and for every $\sigma \in \mathbf{T}^\diamond$, define $f(\sigma) := \min\{k \mid \sigma \in \Sigma_k\}$. It is easy to prove by definition-chasing that $(\mathbf{T}^\diamond, f, 1)$ is a UIRW.

□

## C COMPUTING UNIVERSAL CONSTRAINT PAIRS

In this section, we provide a detailed description of Step 2b of our algorithm, which aims to generate constraint pairs for universal inductive reachability witnesses.

**Step 2b. Computing UIRW Constraint Pairs.** This step is only performed when synthesizing a UIRW and is similar to its IRW variant in Step 2a above. In a UIRW, the universal $\mathbf{T}$-inductive set $\mathbf{T}^\diamond$ should satisfy the condition that for every state $\sigma \in \mathbf{T}^\diamond \setminus \mathbf{T}$, *every* successor $\sigma'$ of $\sigma$ is also in $\mathbf{T}^\diamond$. Moreover, given that $f$ is a universal $\mathbf{T}$−ranking function, we must have $f(\sigma') \leq f(\sigma) - \epsilon$ for every such $\sigma'$.

Let $\ell \in \mathbf{L}$ be a location. The UIRW properties at $\ell$ are equivalent to:

$$\forall \nu \in \mathbb{R}^\mathbf{V}, \ \nu \models \widehat{A_\ell} \Rightarrow \left( \nu \models \tau_\ell \vee \bigwedge_{\theta=(\ell,\ell',\varphi,\mu)} \zeta(\theta) \right) \tag{10}$$

where $\zeta(\theta) = \zeta(\ell, \ell', \varphi, \mu)$ is defined as:

$$\zeta(\theta) := \left( \nu \models \varphi \Rightarrow \left( \mu(\nu) \models \widehat{A_{\ell'}} \wedge \widehat{f_{\ell'}}(\mu(\nu)) \leq \widehat{f_\ell}(\nu) - \widehat{\epsilon} \right) \right) \tag{11}$$

Informally, the constraint in (10) says that if $\nu \models A_\ell$ or equivalently $(\ell, \nu) \in \mathbf{T}^\diamond$, then either $(\ell, \nu) \in \mathbf{T}$, i.e. $\nu \models \tau_\ell$, or *for every transition $\theta$ from $\ell$* the assertion $\xi(\theta)$ holds, i.e. if the transition is possible ($\nu \models \varphi$), then the successor state $(\ell', \mu(\nu))$ is also in $\mathbf{T}^\diamond$, and the $f$ value decreases by at least $\epsilon$ when going to this successor. As in the previous case, the algorithm computes (10) symbolically and writes it in the following equivalent format:

$$\forall \nu \in \mathbb{R}^\mathbf{V}, \left( \nu \models \widehat{A_\ell} \wedge \bigvee_{\theta=(\ell,\ell',\varphi,\mu)} \neg \zeta(\theta) \right) \Rightarrow \nu \models \tau_\ell \tag{12}$$

Let $Q_\ell$ be the LHS assertion above. Similar to Step 2a, $Q_\ell$ is constructed form logical operations and atomic strict/non-strict linear inequalities over $\mathbf{V}$, and its coefficients include the unknown template variables $\widehat{c_{\ell,i,j}}$'s and $\widehat{d_{\ell,j}}$'s defined in Step 1. The algorithm writes $Q_\ell$ in disjunctive normal form, hence obtaining $Q_\ell = Q_{\ell,1} \vee Q_{\ell,2} \vee \ldots \vee Q_{\ell,q}$ in which each $Q_{\ell,i}$ is a conjunction of strict/non-strict linear inequalities over $\mathbf{V}$. It then computes the following constraint pair symbolically:

$$\gamma'_{\ell,i} := (Q_{\ell,i}, \tau_\ell)$$

The algorithm performs these operations for every location $\ell \in \mathbf{L}$ and stores all the resulting $\gamma'_{\ell,i}$ constraint pairs in a set $\Gamma$.

**Example 15.** *In our running example (Figure 2), we are looking for a linear UIRW for the target set $\mathbf{T}' = \{(4, \nu) \mid \nu \models (x \geq y + 4)\}$. In this step, the algorithm creates constraints at every location. We now demonstrate how the process works for location* 3. *At location* 3, *the algorithm considers*

$$\widehat{A_3} \wedge \neg \zeta(\theta_5) \Rightarrow \tau_3$$

*and symbolically computes it as:*

$$\widehat{c_{12}} + \widehat{c_{13}} \cdot x + \widehat{c_{14}} \cdot y \geq 0 \wedge \widehat{c_{15}} + \widehat{c_{16}} \cdot x + \widehat{c_{17}} \cdot y \geq 0 \wedge$$
$$\neg(1 \geq 0 \Rightarrow (\widehat{c_{18}} + 5 \cdot \widehat{c_{19}} + \widehat{c_{19}} \cdot x + \widehat{c_{20}} \cdot y \geq 0 \wedge \widehat{c_{21}} + 5 \cdot \widehat{c_{22}} + \widehat{c_{22}} \cdot x + \widehat{c_{23}} \cdot y \geq 0 \wedge$$
$$\widehat{d_9} + 5 \cdot \widehat{d_{10}} + \widehat{d_{10}} \cdot x + \widehat{d_{11}} \cdot y \leq \widehat{d_6} + \widehat{d_7} \cdot x + \widehat{d_8} \cdot y - \widehat{\epsilon}))$$
$$\Rightarrow (-1 \geq 0)$$

*Note that the transition $\theta_5$ is unconditional, as such we can assume that its condition is simply $1 \geq 0$. Similarly, because there is no target state at location 3, we assume $\tau_3 \equiv (-1 \geq 0)$. Moreover, the transition $\theta_5$ updates the value of $x$ to $x + 5$. This is taken into account when generating the constraint above. The algorithm writes the LHS of the constraint in DNF and handles it exactly as in Example 7.* □

# D PROOFS OF THEOREMS PRESENTED IN SECTION 4

**Notation.** Given a set $X \subseteq \mathbb{R}^V$, we write $\overline{X}$ to denote the closure of $X$, i.e. the smallest closed subset of $\mathbb{R}^V$ that contains $X$. Similarly, for $\Phi$ defined as below, we use the notation $\overline{\Phi}$ to denote the system of linear inequalities obtained by replacing every $\bowtie_i$ in $\Phi$ with $\geq$.

**Corollary 1.** Consider a set $\mathbf{V} = \{v_1, \ldots, v_r\}$ of real-valued variables and the following system of $m$ linear inequalities over $\mathbf{V}$:

$$\Phi : \begin{cases} a_{1,0} + a_{1,1} \cdot v_1 + \ldots + a_{1,r} \cdot v_r \bowtie_1 0 \\ \quad\quad\quad\quad\quad\vdots \\ a_{m,0} + a_{m,1} \cdot v_1 + \ldots + a_{m,r} \cdot v_r \bowtie_m 0 \end{cases}$$

in which $\bowtie_i \in \{>, \geq\}$. When $\Phi$ is satisfiable, it entails a given non-strict linear inequality

$$\psi : c_0 + c_1 v_1 + \ldots + c_r v_r \geq 0$$

if and only if $\psi$ can be written as a non-negative linear combination of $1 \geq 0$ and the inequalities in $\Phi$, i.e. if and only if there exist *non-negative* real numbers $y_0, y_1, \ldots, y_m$, such that:

$$c_0 = y_0 + \sum_{i=1}^{m} y_i \cdot a_{i,0} \quad , \quad c_1 = \sum_{i=1}^{m} y_i \cdot a_{i,1} \quad , \quad \ldots \quad , \quad c_r = \sum_{i=1}^{m} y_i \cdot a_{i,r}.$$

Moreover, $\Phi$ is unsatisfiable if and only if either $-1 \geq 0$ can be derived as above, or $0 > 0$ can be derived as above with the extra requirement that $\sum_{\bowtie_i \in \{>\}} y_i > 0$ (i.e. in order to derive a strict inequality, we should use at least one of the strict inequalities in $\Phi$ with non-zero coefficient).

PROOF. For the first part, suppose that $\psi$ is entailed by $\Phi$, hence $\{x \in \mathbb{R}^V \mid c_0 + c_1 \cdot x_1 + \ldots + c_r \cdot x_r \geq 0\} \supseteq \{x \in \mathbb{R}^V \mid x \models \Phi\}$. The former is a closed set, hence it also includes the closure of the latter, which is the set of points that satisfy $\overline{\Phi}$. Hence, we can apply Lemma 1 to $\overline{\Phi}$ and $\psi$ to obtain the desired result.

For the second part, if $\Phi$ is satisfiable, then obviously no non-negative combination of inequalities in $\Phi$ can sum up to a contradiction such as $0 > 0$ or $-1 \geq 0$. If $\overline{\Phi}$ is not satisfiable, then by Lemma 1, we can obtain $-1 \geq 0$. The only remaining case is if $\overline{\Phi}$ is satisfiable but $\Phi$ is not. Reorder the inequalities in $\Phi$ so that the non-strict inequalities appear first. Then, consider the smallest $i$ for which the first $i$ inequalities in $\Phi$ are unsatisfiable. Let $\Phi[1 \ldots i]$ denote the first $i$ inequalities. Based on our ordering, we know that the $i$-th inequality is strict and of the form $a_{i,0} + a_{i,1} \cdot v_1 + \ldots + a_{i,r} \cdot v_r > 0$. Given that $\Phi[1 \ldots i]$ is unsatisfiable, we know that $\{x \in \mathbb{R}^V \mid x \models \Phi[1 \ldots i-1]\} \subseteq \{x \in \mathbb{R}^V \mid a_{i,0} + a_{i,1} \cdot v_1 + \ldots + a_{i,r} \cdot v_r \leq 0\}$. Therefore, by the first part above, we can write

$$a_{i,0} + a_{i,1} \cdot v_1 + \ldots + a_{i,r} \cdot v_r \leq 0$$

as a non-negative combination of the first $i-1$ inequalities. Moreover, the $i$-th inequality is:

$$a_{i,0} + a_{i,1} \cdot v_1 + \ldots + a_{i,r} \cdot v_r > 0$$

Summing up these two, we get $0 > 0$. □

**Theorem 3** (Soundness). *Given a $k$−linear system $S = (\mathbf{V}, \mathbf{L}, \ell_0, I, \Theta)$, and a $k$−linear set $\mathbf{T}$ of target states, every solution of the non-linear constraint system solved in Step 5 of the algorithm in Section 4.1 produces a valid $k$−linear IRW/UIRW for $\mathbf{T}$ in $S$.*

Proof. Every solution $\mathfrak{s}$ satisfies the constraints generated in Step 3. Therefore, for every constraint pair $\gamma = (\lambda, \varrho) \in \Gamma$ generated in Step 2 and inequality $\alpha_0 + \alpha \cdot \mathbf{V} \geq 0$ in $\varrho$, either $\mathfrak{s}(\lambda)$ is unsatisfiable, i.e. a non-negative linear combination of its inequalities sums up to $0 \geq 1$ or $0 > 0$, or there is such a linear combination that sums up to $\alpha_0 + \alpha \cdot \mathbf{V} \geq 0$. In each case, the coefficients of the combination are given by $\mathfrak{s}(\widehat{y_i})$ for the corresponding $\widehat{y_i}$ variables. Moreover, no matter which case happens, the inequalities in $\varrho$ are entailed by $\lambda$. By definition, the constraint pairs generated in Step 2 modeled inductivity, non-negativity and ranking conditions and hence $\mathfrak{s}$ satisfies these properties. Finally, $\mathfrak{s}$ satisfies the constraints generated in Step 4. Therefore, we have $\mathfrak{s}(\widehat{v_{0,1}}, \ldots, \widehat{v_{0,r}}) \models \mathfrak{s}(\widehat{A_{\ell_0}}) \wedge I$. So, all the requirements for IRW/UIRW are met. □

**Theorem 4** (Completeness). *Given a $k$−linear system $S = (\mathbf{V}, \mathbf{L}, \ell_0, I, \Theta)$, and a $k$−linear set $\mathbf{T}$ of target states, every $k$−linear IRW/UIRW for $\mathbf{T}$ in $S$ is produced by some solution to the non-linear constraint system solved in Step 5 of the algorithm in Section 4.1.*

Proof. We construct the required solution. Let $(\mathbf{T}^\diamond, f, \epsilon)$ be a $k$−linear IRW/UIRW for $\mathbf{T}$ in $S$. Let $A_\ell$ be the set of inequalities defining $\mathbf{T}^\diamond \cap (\ell \times \mathbb{R}^\mathbf{V})$, and $f_\ell$ the linear expression defining $f$ at $\ell$. We use the coefficients in $A_\ell$'s and $f_\ell$'s as the corresponding values for $\mathfrak{s}(\widehat{c_{\ell,i,j}})$'s and $\mathfrak{s}(\widehat{d_{\ell,j}})$'s. Moreover, we let $\mathfrak{s}(\widehat{\epsilon}) = \epsilon$.

By definition, $\mathbf{T}^\diamond$ is an existential/universal $\mathbf{T}$−inductive set, and $f$ is an existential/universal $\mathbf{T}$−ranking function with parameter $\epsilon$. Therefore, $A_\ell$'s and $f_\ell$'s satisfy the constraint pairs generated at Step 2 of the algorithm. By Corollary 1, there are suitable values for each variable $\widehat{y_i}$ such that the constraints in Step 3 are satisfied. We use these values as $\mathfrak{s}(\widehat{y_i})$. Finally, by definition of IRW/UIRW, there exists a valuation $\overline{v} \in \mathbb{R}^\mathbf{V}$ such that $\overline{v} \models A_{\ell_0} \wedge I = \mathfrak{s}(\widehat{A_{\ell_0}}) \wedge I$. We let $\mathfrak{s}(\widehat{v_{0,i}}) = \overline{v}_i$. It is easy to verify that $\mathfrak{s}$ is a solution to the system of non-linear constraints solved in Step 5. □

**Theorem 5** (Complexity). *For fixed constants $k$ and $\beta$, given a $k$−linear $\beta$−branching system $S = (\mathbf{V}, \mathbf{L}, \ell_0, I, \Theta)$, and a $k$−linear set $\mathbf{T}$ of target states, Steps 1–4 of the algorithm in Section 4.1 lead to a polynomial-time reduction from the problem of generating a $k$−linear IRW/UIRW to solving a Quadratic Programming (QP) instance.*

Proof. It is easy to verify that all steps of the algorithm run in polynomial time[¶], and that all the generated (in)equalities over non-program variables are quadratic. However, these (in)equalities are not always combined conjunctively. Specifically, in Step 3, the constraints corresponding to cases (i)–(iii) are combined disjunctively. This being said, we can perform the following actions to obtain a QP instance in polynomial time:

- We first convert every inequality of the form $\mathfrak{e} \bowtie 0$ to $\mathfrak{e} - \widehat{x_\mathfrak{e}} = 0$ by introducing a new variable $\widehat{x_\mathfrak{e}} \bowtie 0$.

---

[¶]The reason for fixing $k$ and $\beta$ is to avoid exponential blow-up when rewriting boolean expressions in DNF.

- We rewrite every disjunction $\mathfrak{e}_1 = 0 \lor \mathfrak{e}_2 = 0$ as $\mathfrak{e}_1 \cdot \mathfrak{e}_2 = 0$. Note that this might create polynomial equalities of higher degree.
- We eliminate terms of degree more than 2 by defining new variables that are equal to their proper divisors, e.g. we rewrite $\widehat{c_1} \cdot \widehat{c_2} \cdot \widehat{c_3}^2$ as $\widehat{v_1} \cdot \widehat{v_2}$ where $\widehat{v_1}, \widehat{v_2}$ are new variables, and add the equalities $\widehat{v_1} = \widehat{c_1} \cdot \widehat{c_2}$ and $\widehat{v_2} = \widehat{c_3}^2$.

The steps above lead to a polynomial blow-up in the size of the system, given that in Step 3 of the algorithm we have disjunctions of at most 3 different boolean formulas. □

**Corollary 2.** Consider a set $\mathbf{V} = \{v_1, \ldots, v_r\}$ of real-valued variables and the following system of $m$ linear inequalities over $\mathbf{V}$:

$$\Phi : \begin{cases} a_{1,0} + a_{1,1} \cdot v_1 + \ldots + a_{1,r} \cdot v_r \bowtie_1 0 \\ \quad\quad\quad\quad\quad \vdots \\ a_{m,0} + a_{m,1} \cdot v_1 + \ldots + a_{m,r} \cdot v_r \bowtie_m 0 \end{cases}$$

in which $\bowtie_i \in \{>, \geq\}$. If $\Phi$ is satisfiable and $\mathrm{Sat}(\Phi)$ is *bounded*, then $\Phi$ entails a given strong polynomial inequality

$$g(v_1, \ldots, v_r) \gg 0,$$

or in other words $\mathrm{Sat}(\Phi) \models g(v_1, \ldots, v_r) \gg 0$, if and only if there exist constants $y_0 \in (0, \infty)$ and $y_1, y_2, \ldots, y_u \in [0, \infty)$, and polynomials $h_1, h_2, \ldots, h_u \in \mathrm{Monoid}(\Phi)$ such that:

$$g = y_0 + \sum_{i=1}^{u} y_i \cdot h_i. \tag{13}$$

PROOF. It is obvious that every $g$ in the form of (13) is strongly positive over $\mathrm{Sat}(\Phi)$, given that $\Phi$ trivially entails $g \geq y_0 > 0$.

We now prove the other side. Suppose $\Phi$ entails $g \gg 0$. Let $\delta > 0$ be the positivity gap of $g$ over $\mathrm{Sat}(\Phi)$ and choose $\delta', y_0$ such that $0 < y_0 < \delta' < \delta$. So, $\mathrm{Sat}(\Phi) \subseteq \mathrm{Sat}(g > \delta')$ and hence $\mathrm{Sat}(\overline{\Phi}) = \overline{\mathrm{Sat}(\Phi)} \subseteq \overline{\mathrm{Sat}(g > \delta')} = \mathrm{Sat}(g \geq \delta')$. Therefore, $\overline{\Phi}$ entails $g - \delta' \geq 0$. So, it also entails $g - y_0 > 0$. Applying Theorem 6 to $\overline{\Phi}$ and $g - y_0$, we have:

$$g - y_0 = \sum_{i=1}^{u} y_i \cdot h_i$$

which is equivalent to Equation (13). □

**Corollary 3.** Consider a set $\mathbf{V} = \{v_1, \ldots, v_r\}$ of real-valued variables and the following system of $m$ polynomial inequalities over $\mathbf{V}$:

$$\Phi : \begin{cases} g_1(v_1, \ldots, v_r) \bowtie_1 0 \\ \quad\quad\quad\quad \vdots \\ g_m(v_1, \ldots, v_r) \bowtie_m 0 \end{cases}$$

in which every $g_i \in \mathbb{R}[\mathbf{V}]$ is a polynomial and every $\bowtie_i \in \{>, \geq\}$. Also, assume that there is some $i$ such that the set $\mathrm{Sat}(g_i \geq 0)$ is compact, or equivalently $\mathrm{Sat}(g_i \bowtie_i 0)$ is bounded. If $\Phi$ is satisfiable, then it entails a *strong* polynomial inequality

$$g(v_1, \ldots, v_r) \gg 0,$$

if and only if there exist a constant $y_0 \in (0, \infty)$ and polynomials $h_0, \ldots, h_m \in \mathbb{R}[\mathbf{V}]$ such that

$$g = y_0 + h_0 + \sum_{i=1}^{m} h_i \cdot g_i \qquad (14)$$

and every $h_i$ is a sum of squares, i.e. $h_i = \sum h_{i,j}^2$ for some polynomials $h_{i,j} \in \mathbb{R}[\mathbf{V}]$.

PROOF. It is obvious that any polynomial $g$ that can be represented as in Equation (14) is strongly positive over $\text{SAT}(\Phi)$ and has a positivity gap of at least $y_0 > 0$.

We now prove the other side. Suppose $\Phi$ is satisfiable and entails $g \gg 0$ with positivity gap $\delta$, and choose $0 < y_0 < \delta' < \delta$. We have $\text{SAT}(\Phi) \subseteq \text{SAT}(g > \delta')$ so $\text{SAT}(\overline{\Phi}) = \overline{\text{SAT}(\Phi)} \subseteq \overline{\text{SAT}(g > \delta')} = \text{SAT}(g \geq \delta') \subseteq \text{SAT}(g > y_0)$. Hence, $\overline{\Phi}$ entails $g - y_0 > 0$. Applying Theorem 10 to $\overline{\Phi}$ and $g - y_0$ leads to the desired result.

□

**Theorem 12.** Consider a set $\mathbf{V} = \{v_1, \ldots, v_r\}$ of real-valued variables and the following system of $m$ polynomial inequalities over $\mathbf{V}$:

$$\Phi : \begin{cases} g_1(v_1, \ldots, v_r) \bowtie_1 0 \\ \quad \vdots \\ g_m(v_1, \ldots, v_r) \bowtie_m 0 \end{cases}$$

in which every $g_i \in \mathbb{R}[\mathbf{V}]$ is a polynomial and every $\bowtie_i \in \{>, \geq\}$. $\Phi$ is unsatisfiable, if and only if at least one of the following statements holds:

(i) There exist a constant $y_0 \in (0, \infty)$ and sum-of-square polynomials $h_0, \ldots, h_m \in \mathbb{R}[\mathbf{V}]$ such that

$$-1 = y_0 + h_0 + \sum_{i=1}^{m} h_i \cdot g_i.$$

(ii) There exist a non-negative integer $\alpha$ and polynomials $h_1, \ldots, h_m \in \mathbb{R}[\mathbf{V}^*]$ for $\mathbf{V}^* = \mathbf{V} \cup \{w_1, \ldots, w_m\}$, such that for some $1 \leq j \leq m$ with $\bowtie_j \in \{>\}$, we have

$$w_j^{2 \cdot \alpha} = \sum_{i=1}^{m} h_i \cdot (g_i - w_i^2)$$

PROOF. If $\Phi$ is satisfiable, then it cannot entail $-1 > 0$, so (i) is impossible. We now show that (ii) implies unsatisfiability of $\Phi$ as well. Define $\tilde{g}_i(v_1, \ldots, v_r, w_1, \ldots, w_m) := g_i(v_1, \ldots, v_r) - w_i^2$. So, we have

$$w_j^{2 \cdot \alpha} = \sum_{i=1}^{m} h_i \cdot \tilde{g}_i.$$

Moreover, $g_j^\alpha = \left(\tilde{g}_j + w_j^2\right)^\alpha = \sum_{i=0}^\alpha \binom{\alpha}{i} \tilde{g}_j^i \cdot w_j^{2\cdot(\alpha-i)} = w_j^{2\cdot\alpha} + h_j' \cdot \tilde{g}_j$ for some $h_j' \in \mathbb{R}[\mathbf{V}^*]$. So, letting $h_i'' = h_i$ for $i \neq j$ and $h_j'' = h_j + h_j'$, we have

$$g_j^\alpha = \sum_{i=1}^m h_i'' \cdot (g_i - w_i^2)$$

Let $v \in \mathbb{R}^\mathbf{V} \cap \text{Sat}(\Phi)$. We extend $v$ to $v^* \in \mathbb{R}^{\mathbf{V}^*}$ as follows: for every $w_i$, let $v^*(w_i) = \sqrt{v(g_i)}$. So, we have $v^*(g_i - w_i^2) = 0$, and hence the RHS of the equation above is 0 at $v^*$. On the other hand, we have $v^*(g_j^\alpha) = v(g_j^\alpha) = (v(g_j))^\alpha > 0$. This contradiction shows that $\Phi$ is unsatisfiable.

We now prove the other side. Suppose that $\Phi$ is unsatisfiable. If $\overline{\Phi}$ is unsatisfiable, then it entails $-1.5 > 0$ and hence we can apply Theorem 10 to write $-1.5 = h_0 + \sum_{i=1}^m h_i \cdot g_i$ for some sum-of-squares polynomials $h_i$, which is equivalent to $-1 = 0.5 + h_0 + \sum_{i=1}^m h_i \cdot g_i$, hence leading to case (i) above. The only remaining case is if $\overline{\Phi}$ is satisfiable but $\Phi$ is not. Reorder the inequalities in $\Phi$ so that the non-strict inequalities appear first. Let $j$ be the smallest index for which $\Phi[1 \ldots j]$, i.e. the set of first $j$ inequalities in $\Phi$, is unsatisfiable. By definition, $\Phi[1 \ldots j-1]$ is satisfiable and hence $\overline{\text{Sat}(\Phi[1 \ldots j-1])} = \text{Sat}(\overline{\Phi}[1 \ldots j-1])$. Moreover, since $\Phi[1 \ldots j] = \Phi[1 \ldots j-1] \wedge (g_j > 0)$ is unsatisfiable, we know that $\Phi[1 \ldots j-1]$ entails $g_j \leq 0$. In other words, $\text{Sat}(\Phi[1 \ldots j-1]) \subseteq \text{Sat}(g_j \leq 0)$. Taking closures from both sides shows that $\overline{\Phi}[1 \ldots j-1]$ entails $g_j \leq 0$. So, $\overline{\Phi}[1 \ldots j]$ entails $g_j = 0$. Define $\tilde{g}_i(v_1, \ldots, v_r, w_1, \ldots, w_m) := g_i(v_1, \ldots, v_r) - w_i^2$. We claim there is no valuation $v^* \in \mathbb{R}^{\mathbf{V}^*}$ such that for all $1 \leq i \leq j$, $\tilde{g}_i(v^*) = 0$, but $g_j(v^*) \neq 0$. To prove this, suppose that such a valuation exists, and let $v$ be its restriction to $\mathbf{V}$. For each $1 \leq i \leq j$, since $\tilde{g}_i(v^*) = 0$, we have $g_i(v) \geq 0$. Moreover, $g_j(v) = g_j(v^*) \neq 0$. This is a contradiction with the previously proven fact that $\overline{\Phi}[1 \ldots j]$ entails $g_j = 0$. Applying the strong nullstellensatz (Theorem 12) to the $\tilde{g}_i$'s and $g_j$, we conclude that there exist a non-negative integer $\alpha$ and polynomials $\tilde{h}_1, \ldots, \tilde{h}_j \in \mathbb{R}[\mathbf{V}^*]$ such that

$$g_j^\alpha = \sum_{i=1}^j \tilde{h}_i \cdot \tilde{g}_i$$

Note that $g_j^\alpha = \left(\tilde{g}_j + w_j^2\right)^\alpha = \sum_{i=0}^\alpha \binom{\alpha}{i} \tilde{g}_j^i \cdot w_j^{2\cdot(\alpha-i)} = w_j^{2\cdot\alpha} + h_j' \cdot \tilde{g}_j$ for some $h_j' \in \mathbb{R}[\mathbf{V}^*]$. Defining $h_i = \tilde{h}_i$ for all $i \neq j$, and $h_j = \tilde{h}_j - h_j'$, we get

$$w_j^{2\cdot\alpha} = \sum_{i=1}^j h_i \cdot \tilde{g}_i = \sum_{i=1}^j h_i \cdot (g_i - w_i^2).$$

□

## E ENFORCING A POLYNOMIAL TO BE A SUM OF SQUARES

In several places in our algorithm, we have a sum-of-squares polynomial $\widehat{h}$ defined by a template

$$\widehat{h} := \widehat{\eta_1} \cdot \mathfrak{m}_1 + \ldots + \widehat{\eta_\mathfrak{n}} \cdot \mathfrak{m}_\mathfrak{n}$$

in which $\{\mathfrak{m}_1, \ldots, \mathfrak{m}_\mathfrak{n}\}$ are monomials over a set $\mathbf{V}$ of variables, the $\widehat{\eta_i}$'s are unknown reals, and the algorithm depends on ensuring that $\widehat{h}$ is indeed a sum-of-squares polynomial. Given that our algorithm reduces the problem of generating an IRW/UIRW to quadratic programming, we would like to similarly reduce the problem of $\widehat{h}$ being a sum-of-squares to quadratic programming over the $\widehat{\eta_i}$'s. In this section, we show how such a reduction can be obtained. This is a standard procedure and has previously been used in many other constraint-based program analysis algorithms. The presentation we use is taken from [Chatterjee et al. 2020]. Our main tools are two well-known theorems:

**Theorem 16** ([Horn and Johnson 1990, Chapter 7]). *A polynomial $\widehat{h} \in \mathbb{R}[\mathbf{V}]$ of even degree $\mathfrak{d}$ is a sum-of-squares if and only if there exists an $\mathfrak{r}$−dimensional symmetric positive semi-definite matrix $\mathcal{P}$ such that $h = y^T \mathcal{P} y$, where $\mathfrak{r}$ is the number of monomials of degree no greater than $\mathfrak{d}/2$ and $y$ is a column vector consisting of every such monomial.*

**Theorem 17** ([Higham 2009]). *A symmetric square matrix $\mathcal{P}$ is positive semi-definite if and only if it has a Cholesky decomposition of the form $\mathcal{P} = \mathcal{L}\mathcal{L}^T$ where $\mathcal{L}$ is a lower-triangular matrix with non-negative diagonal entries.*

Given the two theorems above, we use the following standard process for generating quadratic equations that are equivalent to $\widehat{h}$ being a sum-of-squares:

***Generating Sum-of-Squares Constraints.*** The algorithm generates the set $M_{\frac{\mathfrak{d}}{2}}$ consisting of all monomials of degree at most $\frac{\mathfrak{d}}{2}$ over $\mathbf{V}$ and creates a vector $y$ of these monomials. It then symbolically computes the following equality:

$$\widehat{h} = y^T \widehat{\mathcal{L}} \widehat{\mathcal{L}}^T y.$$

Here, $\widehat{\mathcal{L}}$ is a lower-triangular matrix. Every entry of $\widehat{\mathcal{L}}$ is a new unknown variable, and every diagonal entry is constrained to be non-negative. As usual, the algorithm equates the corresponding terms on both sides of this polynomial equality to obtain quadratic equations over the unknown variables. It follows directly from the two theorems above that this reduction is both sound and complete.

**Example 16** (Taken from [Chatterjee et al. 2020]). *Let $\mathbf{V} = \{a, b\}$ be the set of variables and $\widehat{h} \in \mathbb{R}[\mathbf{V}]$ a quadratic polynomial, i.e. $\widehat{h}(a, b) = \widehat{\eta_1} + \widehat{\eta_2} \cdot a + \widehat{\eta_3} \cdot b + \widehat{\eta_4} \cdot a^2 + \widehat{\eta_5} \cdot a \cdot b + \widehat{\eta_6} \cdot b^2$. We aim to encode the property that $\widehat{h}$ is a sum-of-squares as a system of quadratic equalities and inequalities. To do so, we first generate all monomials of degree at most $\lfloor \mathfrak{d}/2 \rfloor = 1$, which are 1, $a$ and $b$. Hence, we let $y = \begin{bmatrix} 1 & a & b \end{bmatrix}^T$. We then generate a template for a lower-triangular matrix $\widehat{\mathcal{L}}$ whose every non-zero entry is a new variable:*

$$\widehat{\mathcal{L}} = \begin{bmatrix} \widehat{l_1} & 0 & 0 \\ \widehat{l_2} & \widehat{l_3} & 0 \\ \widehat{l_4} & \widehat{l_5} & \widehat{l_6} \end{bmatrix}.$$

*We also add the inequalities $\widehat{l_1} \geq 0, \widehat{l_3} \geq 0$ and $\widehat{l_6} \geq 0$ to our system. Now, we write the equation $\widehat{h} = y^T \widehat{\mathcal{L}} \widehat{\mathcal{L}}^T y$ and compute it symbolically:*

$$\widehat{h} = \begin{bmatrix} 1 & a & b \end{bmatrix} \begin{bmatrix} \widehat{l_1} & 0 & 0 \\ \widehat{l_2} & \widehat{l_3} & 0 \\ \widehat{l_4} & \widehat{l_5} & \widehat{l_6} \end{bmatrix} \begin{bmatrix} \widehat{l_1} & \widehat{l_2} & \widehat{l_4} \\ 0 & \widehat{l_3} & \widehat{l_5} \\ 0 & 0 & \widehat{l_6} \end{bmatrix} \begin{bmatrix} 1 \\ a \\ b \end{bmatrix},$$

which leads to:

$$\widehat{\eta_1} + \widehat{\eta_2} \cdot a + \widehat{\eta_3} \cdot b + \widehat{\eta_4} \cdot a^2 + \widehat{\eta_5} \cdot a \cdot b + \widehat{\eta_6} \cdot b^2 =$$
$$\widehat{l_1}^2 + 2 \cdot \widehat{l_1} \cdot \widehat{l_2} \cdot a + 2 \cdot \widehat{l_1} \cdot \widehat{l_4} \cdot b + (\widehat{l_2}^2 + \widehat{l_3}^2) \cdot a^2 + (2 \cdot \widehat{l_2} \cdot \widehat{l_4} + 2 \cdot \widehat{l_3} \cdot \widehat{l_5}) \cdot a \cdot b + (\widehat{l_4}^2 + \widehat{l_5}^2 + \widehat{l_6}^2) \cdot b^2.$$

Note that both sides of the equation above are polynomials over $\{a, b\}$, hence they are equal iff their corresponding coefficients are equal. So, we get the following quadratic equalities over the $\widehat{\eta_i}$'s and $\widehat{l_i}$'s: $\widehat{\eta_1} = \widehat{l_1}^2, \widehat{\eta_2} = 2 \cdot \widehat{l_1} \cdot \widehat{l_2}, \ldots, \widehat{\eta_6} = \widehat{l_4}^2 + \widehat{l_5}^2 + \widehat{l_6}^2$. This concludes the construction of our quadratic system. □

## F POLYNOMIAL PROGRAMS USED IN THE EXPERIMENTAL RESULTS

In this section, we provide details of the polynomial programs that were used in our experiments in Section 5. Figures 5–10 show the programs. We now discuss each benchmark in more detail:

- sqrt1: This program gets a value $n$ as input and computes $\lfloor \sqrt{n} \rfloor$ by simply iterating through every integer starting from 1. The goal is to reach the end of the program with $n - s > 10^5$. Therefore, to solve this task, a verifier has to assign a value to $n$ such that $n - \lfloor \sqrt{n} \rfloor > 10^5$. It is easy to see that any $n > 10^5 + 316$ works. However, this example is interesting because the shortest path to a target state needs to go through 316 iterations of the **while** loop. Moreover, the loop has a quadratic guard. As such, a verifier that is based on abstract interpretation needs to obtain a relatively fine abstraction, whereas approaches based on loop-unrolling and symbolic execution need to unroll this non-linear loop 316 times. As mentioned in Table 2, CPAchecker times out on this instance. However, VeriAbs succeeds in proving reachability in 207.3 seconds. In contrast, our approach synthesizes an IRW in just 40.8 seconds.

$$I : n \geq 1 \ \land \ s = 1$$
$$\textbf{while} \ \ (s + 1)^2 \leq n :$$
$$s := s + 1$$

Fig. 5. The program sqrt1. Our target is to reach the end of this program with $n - s > 10^5$.

- sqrt2: This is a variant of sqrt1 in which the value of $s$ is doubled in every step if $2 \cdot s \leq \sqrt{n}$. This simple change means that there are now many short paths that reach the target. For example, by setting $n = 2^{18}$, one can reach the target in just 9 iterations. Unsurprisingly, both CPAchecker and VeriAbs can handle this example (Table 2). This being said, note that the complexity of our approach does not depend on the length of paths. As such, while sqrt1 is much harder than sqrt2 for other approaches, our runtimes on these two benchmarks are very close. Indeed, our method solves sqrt1 a bit faster than sqrt2 (40.8s vs 46.4s). This is because sqrt1 is a smaller program.

$$I : n \geq 1 \ \wedge \ s = 1$$
$$\textbf{while} \ \ (s+1)^2 \leq n :$$
$$\quad \textbf{if} \ \ 4 \cdot s^2 \leq n :$$
$$\quad\quad s := 2 \cdot s$$
$$\quad \textbf{else} :$$
$$\quad\quad s := s + 1$$

Fig. 6. The program sqrt2. Our target is to reach the end of this program with $n - s > 10^5$.

- sum: This program simply sums up all the integers from 1 to a given value $n$. Note that the program itself is linear (the loop guard and the updates are linear). However, we need polynomial arguments given that for every integer $n$, at the end of this program we will have $s = \frac{n \cdot (n+1)}{2}$. As in the previous examples, the target set can only be reached after many iterations. To reach the target, it suffices to choose $10000 \leq n \leq 11000$. Our algorithm is exact and can handle tight inequalities. We chose this liberal interval in order to make the instance solvable for abstract interpretation approaches with good precision. Nevertheless, CPAchecker timed out and VeriAbs terminated with no result, i.e. returned "unknown".

$$I : s = 0 \ \wedge \ i = 1$$
$$\textbf{while} \ \ i \leq n :$$
$$\quad (s, i) := (s + i, i + 1)$$

Fig. 7. The program sum. Our target is to reach the end of this program with $50005000 \leq s \leq 60505500$.

- sum2: This program is similar to sum but it adds the squares of integers from 1 to $n$. Because this program has non-linear assignments, it is intuitively harder to verify in comparison with sum.

$$I : s = 0 \ \wedge \ i = 1$$
$$\textbf{while} \ \ i \leq n :$$
$$\quad (s, i) := (s + i^2, i + 1)$$

Fig. 8. The program sum2. Our target is to reach the end of this program with $333383335000 \leq s \leq 443727168500$.

- robot1: This program models the behavior of two robots on a 2d plane. One robot is located at $(x_1, y_1)$ and the other at $(x_2, y_2)$. Initially, we have $(x_1, y_1) = (x_2, y_2)$. At each iteration, each robot moves one unit upwards or to the right. The direction is chosen non-deterministically. The goal is to prove reachability to the endpoint of the program. This is equivalent to proving that it is possible for the robots to move in such a way that makes their distance from each other more than $\sqrt{10^5}$. The main difficulty in this program is the combinatorial explosion in the number of paths. Nevertheless, note that a relatively large proportion of the paths lead to the desired target. As such, it was surprising for us to see that our approach was the only one that succeeded in handling this example.

$$I : x_1 = x_2 \ \land \ y_1 = y_2$$
$$\textbf{while} \ \ (x_1 - x_2)^2 + (y_1 - y_2)^2 \leq 10^5 :$$
$$x_1 := x_1 + 1 \ \square \ y_1 := y_1 + 1$$
$$x_2 := x_2 + 1 \ \square \ y_2 := y_2 + 1$$

Fig. 9. The program robot1. Our target is to reach the end of this program.

- robot2: This is a variant of robot1 which intuitively seems to be a bit harder. The same two robots are put on opposite sides of a square with side-length $10^4$ and the goal is to prove that they can move in a way that decreases their distance to less than 10. This example creates the same combinatorial explosion in the number of possible paths as in robot1, but this time, only a very small proportion of these paths reach the target. Nevertheless, our approach can handle this example in virtually the same amount of time as robot1. This is because our approach is (semi-)complete and finds a polynomial IRW if one exists. It does not depend on the proportion of paths that lead to a target state.

$$I : y_1 = y_2 + 10^4 \ \land \ x_1 = x_2 - 10^4$$
$$\textbf{while} \ \ (x_1 - x_2)^2 + (y_1 - y_2)^2 \geq 100 :$$
$$x_1 := x_1 + 1 \ \square \ y_1 := y_1 + 1$$
$$x_2 := x_2 + 1 \ \square \ y_2 := y_2 + 1$$

Fig. 10. The program robot2. Our target is to reach the end of this program.